\documentclass[aps,prb,reprint,superscriptaddress]{revtex4-2}

\usepackage{tikz}
 \normalsize
\usepackage{scalerel}
\usepackage{amssymb}
\usepackage{suffix}
\usepackage{float}
\usepackage{mathtools}
\usepackage[utf8]{inputenc}
\usepackage{booktabs}
\usepackage{cases}
\usepackage[multiple]{footmisc}
\usepackage{dcolumn}
\usepackage{color,soul}
\usepackage{rotating}
\usepackage{perpage}
\usepackage{xcolor}
\usepackage{comment}
\usepackage{soul}
\usepackage{tikz}
\usepackage[T1]{fontenc}
\usepackage{etoolbox}
\usepackage{graphics}
\usepackage{siunitx}
\usepackage{hyperref}
\hypersetup{colorlinks}
\usepackage{float}	
\usepackage{collref}
\usepackage{multirow}
\usepackage{mathtools}
\usepackage{bm}
\usepackage{url}
\usepackage{wasysym}
\definecolor{WildStrawberry}{RGB}{255,67,164}

\begin{document}

\title{Spectral statistics and energy-gap scaling in $k-$local spin Hamiltonians}
\author{Sasanka Dowarah}
\email{sasanka.dowarah@utdallas.edu}
 \affiliation{Department of Physics, The University of Texas at Dallas, Richardson, Texas 75080, USA}
\date{\today}
\begin{abstract}
We investigate the spectral properties of all-to-all interacting spin Hamiltonians acting on exactly $k$ spins, whose coupling coefficients are drawn from a normal distribution with mean $\mu$ and variance $\sigma^2$. For $\mu = 0$, we demonstrate that their associated random matrix ensemble -- Gaussian Orthogonal Ensemble (GOE), Gaussian Unitary Ensemble (GUE), or Gaussian Symplectic Ensemble (GSE) -- is determined by the parity of system size $L$ and locality $k$, following standard time-reversal symmetry classification. For couplings with a nonzero mean, we map the Hamiltonians to deformed random matrix ensembles and analyze conditions for an energy gap between the ground state and the first excited state. For $\mu < 0$, we find two distinct regimes: for $k \gg \sqrt{L}$, the gap closes at critical disorder $\sigma_{c} \approx |\mu|$. Near this transition the energy gap $\Delta$ exhibits universal quadratic scaling $\Delta /L \sim (\sigma - \sigma_{c})^{2}$. When $k \ll \sqrt{L}$, $\sigma_{c}$ scales  with $|\mu|$, but lacks a sharp transition. Our work introduces a semi-solvable model that captures universal features of random-matrix statistics, and spectral gap formation, providing a foundation for systematic extensions to more general many-body systems.
\end{abstract}
\maketitle
{\allowdisplaybreaks}
 \parskip 0 pt

\section{Introduction}
The behavior of complex quantum systems is often shaped by the nature of their interactions and symmetries.  A key challenge in quantum many-body physics is understanding how the complexity and chaotic behavior of quantum systems arise from the underlying interactions and symmetries. The emergence of random-matrix theory statistics in the energy level of a system are considered as a signature that the system is quantum chaotic \cite{PhysRevLett.52.1, PhysRevB.47.14291, DPoilblanc_1993, PhysRevB.75.155111, PhysRevB.82.174411, PhysRevB.91.081103, GUHR1998189}. In this context, Random Matrix Theory (RMT) has become a central tool for analyzing spectral fluctuations and identifying signatures of ergodicity. 
Wigner initially introduced RMT to model the level statistics of heavy atomic nuclei \cite{c9eb8278-b5e2-37bc-a322-8e81785f98ed} and was later extended by Dyson \cite{10.1063/1.1703863, 10.1063/1.1703773}, who systematically classified random matrix ensembles based on symmetry considerations:  $(1)$ time-reversal invariant systems with rotational symmetry for which the Hamiltonian can be made real, $(2)$ systems without time-reversal invariance for which the Hamiltonian is complex Hermitian and $(3)$ time-reversal invariant systems with half-integer spin and broken rotational symmetry where the Hamiltonian can be written in terms of quaternions. These three classifications define three universal ensembles: the Gaussian Orthogonal Ensemble (GOE), the Gaussian Unitary Ensemble (GUE), and the Gaussian Symplectic Ensemble (GSE). The strength of random matrix theory lies in its universality: the statistical properties of a random matrix ensemble do not depend on the specific details of the matrix elements, but rather only on the fundamental symmetry of the system. This universality has been observed across a wide range of quantum chaotic systems, where the spectral statistics exhibit remarkable consistency regardless of the specific form of the Hamiltonian.

A spectral gap refers to the nonzero energy difference between the ground state and the first excited state in the thermodynamic limit. The presence or absence of such a gap is a central concept in quantum many-body physics as it governs the nature of low-energy excitations, stability, and dynamical behavior of the phase -- gapped Hamiltonians exhibit exponentially decaying correlations with distance \cite{Hastings2006, Hastings_2007, Nachtergaele2006, Arad2017}, whereas gapless Hamiltonians can display long-range correlations, and the closing of the gap often signals a quantum phase transition \cite{10.1093/acprof:oso/9780199227259.001.0001, RevModPhys.69.315, Bachmann2012}. Determining whether a given local Hamiltonian is gapped or gapless is, in general, undecidable -- there exists no algorithm capable of determining the spectral gap for all possible Hamiltonians \cite{Cubitt2015, PhysRevX.10.031038}. Nevertheless, for many important classes of models, rigorous results are known. For example, the Haldane spin-1 chain, the transverse-field Ising model, and the XXZ chain in certain parameter regimes are provably gapped \cite{PhysRevB.69.104431, Hastings2006, Hastings_2007, KOMA1997}, while others such as the Heisenberg spin-1/2 chain and critical Ising models are rigorously gapless \cite{LiebMattis1966, PFEUTY197079}.

For all-to-all interacting Hamiltonians, the system is effectively zero-dimensional, meaning there is no notion of spatial distance as in local $1$D or $2$D models. However, the spectral gap has an effect on long-ranged and all-to-all interacting models in the stability of nonthermal states (quasi-stationary states) \cite{doi:10.1073/pnas.2101785118}, behavior of correlation function in time in the spin-glass models, and other low energy behavior. Therefore, it is essential to explore the spectral gap in long-range and all-to-all interacting systems as a foundation for characterizing their dynamical and thermodynamic properties.

In this work, we study all-to-all interacting Hamiltonians consisting of terms that act on exactly $k$ spins. The couplings are drawn from a normal distribution with mean $\mu$ and variance $\sigma^{2}$. We first demonstrate that the universality in spectral statistics for the model with $\mu = 0$ depends solely on the system size and the locality of interactions: classification of the Hamiltonian into GOE, GUE, or GSE is entirely determined by whether these parameters are even or odd. Additionally, we provide numerical evidence to support our results.  A similar parity-based classification scheme exists for SYK models with Majorana and complex fermions \cite{PhysRevD.94.126010, Kanazawa2017, PhysRevLett.124.244101, PhysRevX.12.021040, PhysRevB.95.115150}.

 We show that when the couplings in the Hamiltonians are chosen from a normal distribution with nonzero mean, then these Hamiltonians can be mapped to \emph{deformed random matrices} with the random matrix part being GOE/GUE/GSE depending on the parity of $L$ and $k$. This enables us to determine the critical value of the variance $\sigma^{2}$ for which the Hamiltonian has an energy gap between the ground state and the first excited state. For $\sigma < 0$, we find that for $k \gg \sqrt{L}$, the $k-$local spin Hamiltonian has an energy gap as long as $\sigma < |\mu|$. For $k \ll \sqrt{L}$, we show that the critical disorder $\sigma_{c}$, at which the energy gap closes, scales proportionally with $\mu$, although this crossover lacks a sharp cutoff. We also derive the scaling of the gap near the critical point in the limit $k \gg \sqrt{L}$.

The rest of the paper is organized as follows: in Section \ref{sec:model-and-definitions}, we introduce the $k-$local Hamiltonian model, which is the main focus of our analysis. The results section is divided into two parts. In Part \hyperref[subsec:part_I]{I}, we present a classification of these models based on their system size and their locality being odd or even. In Part \hyperref[subsec:energy_gap]{II}, we present an analysis of the energy gap in these Hamiltonians when the mean of the couplings is nonzero.
\vspace{5 mm}
\section{Model and definitions}\label{sec:model-and-definitions}
The set of all tensor products of single-qubit Pauli matrices and the identity forms a basis for the $2^{L} \times 2^{L}$ Hermitian matrices. For a $L$ qubit system this basis has $4^{L}$ elements of the form $\sigma_{\alpha_{1}} \otimes \sigma_{\alpha_{2}} \otimes \cdots \otimes \sigma_{\alpha_{L}}$, with $\sigma_{\alpha_{j}} \in \{ I, \sigma_{x}, \sigma_{y}, \sigma_{z}\}$, and is orthogonal with respect to the inner product $\mathrm{Tr}[A^{\dagger} B] = 2^{L} \delta_{AB}$. Any Hermitian matrix -- thus any Hamiltonian $H$ can be uniquely represented as
\begin{eqnarray}
    H &=& \sum_{j} c_{j} \bigotimes^{L}_{l=1} \sigma_{\alpha_{l}}, \quad c_{j} \in \mathbb{R}.
\end{eqnarray}
A Hamiltonian $H$ acting on a system of $L$ qubits is defined as $k-$local if it can be expressed as a sum of terms $H=\sum_{j}H_{j}$, where each term $H_{j}$ is a Hermitian operator that acts nontrivially on at most $k$ particles or spins \cite{annurev:/content/journals/10.1146/annurev-conmatphys-031214-014726, doi:10.1137/S0097539704445226}. In this work, we define the following Hamiltonian that acts on exactly $k$ spins:
\begin{widetext}
\begin{equation}
H(\mu, \sigma^{2}) \equiv \sum_{j_{1}< j_{2} < \cdots < j_{k}} \sum^{3}_{\alpha_{1}, \cdots, \alpha_{k} = 1}  J^{\alpha_{1} \alpha_{2} \cdots \alpha_{k}}_{j_{1}j_{2}\cdots j_{k}}
   \; \sigma^{(j_{1})}_{\alpha_{1}}\otimes\sigma^{(j_{2})}_{\alpha_{2}}\otimes
    \cdots \otimes\sigma^{(j_{k})}_{\alpha_{k}}\;, \label{eq:k_local_Hamiltonian_model}
\end{equation}
\end{widetext}
where $\sigma^{(j)}_{\alpha}$ represents a Pauli matrix $\sigma_{\alpha} \in \{\sigma_{x}, \sigma_{y}, \sigma_{z}\}$ acting on the $j$th qubit, and $J^{\alpha_{1} \alpha_{2} \cdots \alpha_{k}}_{j_{1}j_{2}\cdots j_{k}}$ are real numbers from the normal distribution with mean and variance \footnote{We note that the variance in Eq.~\eqref{eq:Mean_variance_of_couplings_Hamiltonian} scales as $L^{2}/ \binom{L}{k}$, which differs from the standard SYK or $p$-spin model convention. In the regime $k \gg \sqrt{L}$, the spectrum of $H(0, 1)$ follows a Wigner semicircle distribution with a sharp edge and no tails. Since the eigenvalues of the deterministic term $H(1, 0)$ are known exactly (see Eq.~\eqref{eq:analytical_eigenvalues_of_disorder_free_Hamiltonian}), we choose a normalization to scale its spectrum as $O(L)$. This fixes the normalization of the mean in Eq.~\eqref{eq:Mean_variance_of_couplings_Hamiltonian}. To ensure that the disordered part remains competitive in the thermodynamic limit, we must subsequently choose the normalization of the disordered term such that it also scales as $O(L)$. This requires us to choose the variance of the couplings as $L^{2}/\binom{L}{k}$.}
\begin{eqnarray}
    \text{Mean}(J^{\alpha_{1} \alpha_{2} \cdots \alpha_{k}}_{j_{1}j_{2}\cdots j_{k}}) = \frac{\mu L}{\sqrt{3^{k}} \binom{L}{k}}, \; 
    \text{Var}(J^{\alpha_{1} \alpha_{2} \cdots \alpha_{k}}_{j_{1}j_{2}\cdots j_{k}}) =  \frac{\sigma^{2} L^{2}}{3^{k} \binom{L}{k}},\; \label{eq:Mean_variance_of_couplings_Hamiltonian}
\end{eqnarray}
where the normalizations are chosen such that the energy of the Hamiltonian grows linearly with system size, and comparisons are meaningful across different system sizes, following Kac prescription \cite{10.1063/1.1703946, kastner2025longrangesystemsnonextensivityrescaling, RevModPhys.95.035002}. All-to-all interacting models like Eq.\eqref{eq:k_local_Hamiltonian_model} where every spin couples to every other spin, are effectively zero-dimensional and lack any structure. Various properties of these models in the $\mu = 0$ limit have been studied under different names -- quantum $q-$spin model \cite{swingle2023bosonicmodelquantumholography, Xu_2025}, quantum $p-$spin glass  \cite{Erdős2014, PhysRevX.10.031026}, SpinXYq  \cite{Hanada2024}, random energy model \cite{PhysRevLett.45.79} etc. This list is not exhaustive, and we will not discuss the results of these studies here as they are not directly relevant to the main focus of this work.

We now show that the random matrix ensemble of these Hamiltonians depends only on whether the system size and the locality of interaction are even or odd. To this end, we first determine the random matrix ensemble of the Hamiltonians numerically, using the energy level statistics, which is defined as the gap ratio of consecutive eigenvalues  \cite{PhysRevB.75.155111, PhysRevLett.110.084101}
\begin{eqnarray}
    r_n = \frac{\min(\delta_{n}, \delta_{n-1})}{\max(\delta_{n}, \delta_{n-1})} = \min\left(\tilde{r}_n, \frac{1}{\tilde{r}_n}\right), \label{eq:energy_level_statistics_definition}
\end{eqnarray}
where
\begin{eqnarray}
\tilde{r}_{n} = \frac{\delta_{n}}{\delta_{n-1}},
\end{eqnarray}
and $\delta_{n} = E_{n+1} - E_{n}$, is the nearest-neighbor spacing in the ordered energies $E_{n}$. This quantity has the advantage that it does not require unfolding of the energy spectrum, which requires knowledge of the density of states of the system \cite{Haake2018QuantumSignatures}. In the ergodic (chaotic) regime, the average value of $r_{n}$ approaches the universal values predicted by random matrix theory: $0.53$ for GOE, $0.60$ for GUE, and $0.67$ for GSE  \cite{PhysRevLett.110.084101}. In contrast, for integrable or many-body localized systems, the statistics follow a Poisson distribution, yielding an average $\langle r_{n}\rangle \approx 0.386$. Thus, $r_{n}$ serves as a powerful and efficient diagnostic for distinguishing chaotic from regular spectral behavior in many-body quantum systems. Note that one needs to remove all unwanted symmetries from the Hamiltonian before calculating the $r_{n}$ using Eq. \eqref{eq:energy_level_statistics_definition} \cite{PhysRevX.12.011006}. \par
Another indicator of ergodicity is the repulsion between the nearby energy levels. To quantify this, one needs to perform the unfolding of the eigenvalue spectrum \cite{Haake2018QuantumSignatures}. The unfolding procedure maps the raw spectrum $\{E_{n}\}$ to a new set in which the mean level density is unity. From standard random matrix theory results, the analytical expressions for these probability distributions of this spacing are given by \cite{Mehta2004}
\begin{eqnarray}
    p_{\mathrm{GOE}}(s) &=& (\pi / 2) s e^{-\pi s^{2}/4}, \\
    p_{\mathrm{GUE}}(s) &=& (32 / \pi^{2}) s^{2} e^{ - 4 s^{2} / \pi}, \\
    p_{\mathrm{GSE}}(s) &=& (2^{18} /(3^{6}\pi^{3})) s^{4} e^{-64 s^{2} / (9\pi)}. \label{eq:analytical_normalized_spacing}
\end{eqnarray}

For $\mu \neq 0$ and $\sigma=0$, all the coupling coefficients are identical: $J^{\alpha_{1} \alpha_{2}\cdots j_{k}}_{j_{1} j_{2} \cdots j_{k}} \equiv \frac{\mu L}{\sqrt{3^{k}} \binom{L}{k}}$. As a result, the Hamiltonian in this case has degenerate eigenvalues due to many additional symmetries (see Sec. \ref{subsec:eigenstructure_in_the_absence_of_disorder} for details). In contrast, when $\mu=0$ and $\sigma > 0$, the couplings are fully disordered, and the randomness breaks these symmetries, and the level statistics of the Hamiltonian correspond to one of the random matrix ensembles. To ensure that we analyze the Hamiltonian without any unwanted symmetry, we start by rearranging Eq. \eqref{eq:k_local_Hamiltonian_model}:
\begin{widetext}
\begin{eqnarray}
    H (\mu, \sigma^{2})
    &=& \sum_{j_{1}< j_{2} < \cdots < j_{k}}  \sum^{3}_{\alpha_{1}, \cdots, \alpha_{k} = 1} \bigg[ \frac{\mu L}{\sqrt{3^{k}} \binom{L}{k}} \sigma^{(j_{1})}_{\alpha_{1}}\sigma^{(j_{2})}_{\alpha_{2}}
    \cdots \sigma^{(j_{k})}_{\alpha_{k}} + \frac{\sigma L}{\sqrt{3^{k} \binom{L}{k}}} \frac{J^{\alpha_{1} \alpha_{2} \cdots \alpha_{k}}_{j_{1}j_{2}\cdots j_{k}}- \frac{\mu L}{\sqrt{3^{k}} \binom{L}{k}}}{\frac{\sigma L}{\sqrt{3^{k} \binom{L}{k}}}} \; \sigma^{(j_{1})}_{\alpha_{1}}\sigma^{(j_{2})}_{\alpha_{2}}
    \cdots \sigma^{(j_{k})}_{\alpha_{k}} \bigg],\; \\
    &\equiv& \frac{\mu L}{\sqrt{3^{k}} \binom{L}{k}} H(1,0) + \frac{\sigma L}{\sqrt{3^{k} \binom{L}{k}}} H(0,1), \label{eq:H-mu-sigma-decomposition}
\end{eqnarray}
\end{widetext}
where \(H(1,0)\) represents the fixed, deterministic part of the Hamiltonian with all coupling terms set to unity, and \(H(0,1)\) corresponds to the completely disordered part. The decomposition in Eq.\eqref{eq:H-mu-sigma-decomposition} isolates the effects of randomness from structured interactions. In our numerical study, we focus on the regime \( \mu = 0 \) and \( \sigma = 1 \), thereby analyzing the level statistics of \( H(0,1) \) as a representative of the fully random ensemble. Since \( H(0, \sigma^{2})=\sigma H(0,1) \), the spectral statistics of the two are equivalent to an overall scaling. Focusing on the level statistics of this completely disordered Hamiltonian $H(0,1)$ ensures that we do not encounter any accidental symmetries.

\section{Results}\label{sec:results}
\subsection*{Part I: Classification} \label{subsec:part_I}
\begin{table*}[t]
\caption{Dependence of energy level statistics of the $k-$local Hamiltonians on system size $L$ and locality $k$.}
\label{tab:yourlabel}
\begin{ruledtabular}
\begin{tabular*}{\textwidth}{@{\extracolsep{\fill}} lcccc}
\textrm{System Size} & \textrm{Locality} & \textrm{RM ensemble} & \textrm{$T^{2}$} & 
\textrm{Commutation relation}\\
\hline
Odd & Even & GSE & -1 & $[H, T] = 0$\\
Even & Even & GOE & 1 & $[H, T] = 0$\\
Even & Odd & GUE & $\times$ & $\{H, T \} = 0$\\
Odd & Odd & GUE & $\times$ & $\{H, T \} = 0$\\
\end{tabular*}
\end{ruledtabular}
\end{table*}
\begin{figure}
    \centering
    \includegraphics[width=\linewidth]{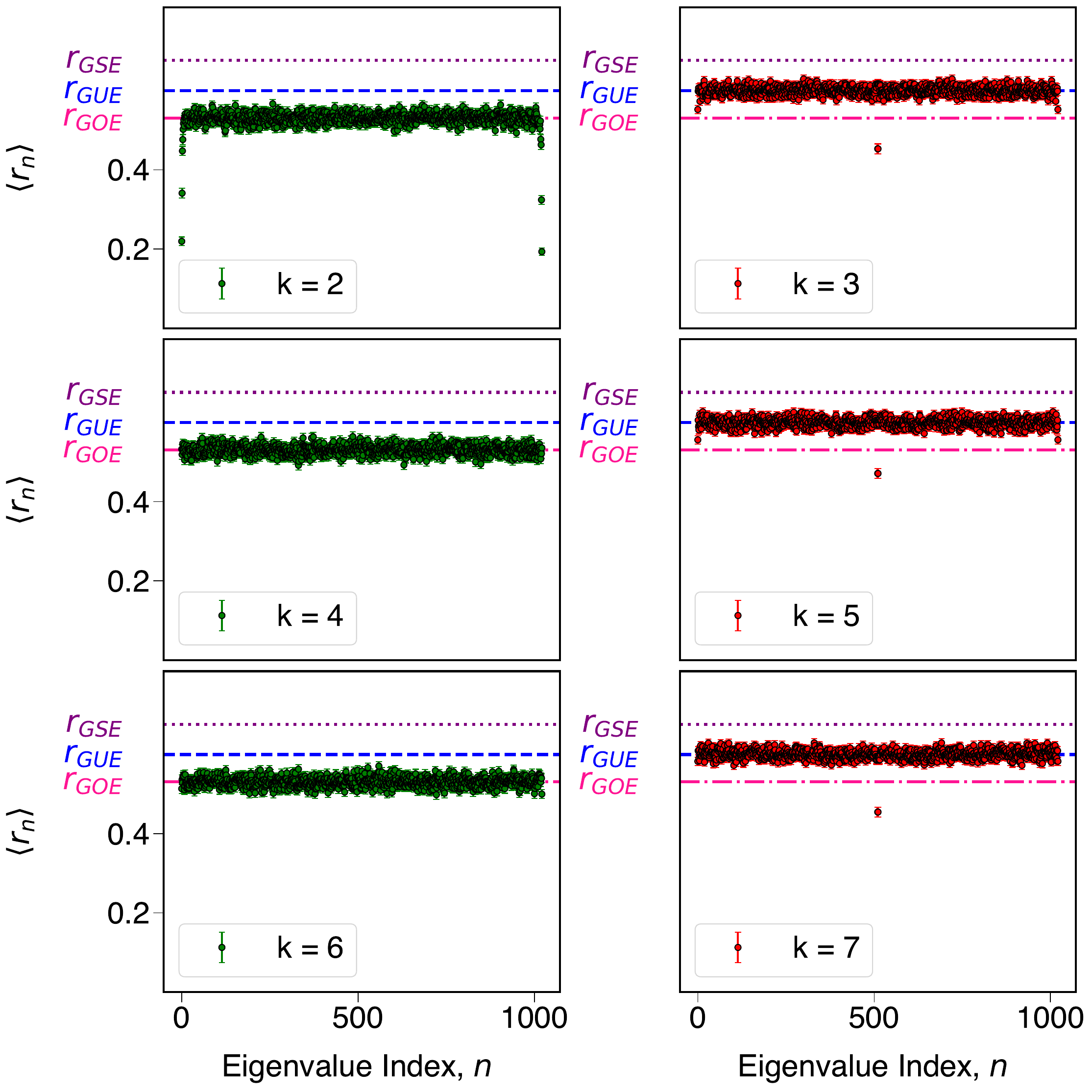}
    \caption{Average level statistics $\langle r_{n} \rangle$ of the Hamiltonian $H(0,1)$ for $L=10$ for locality $k=2,3,4,5,6,7$. The error bar indicates the standard error of the mean for $480$ different random realizations of the couplings. The pink dashed-dotted line, the blue dashed line, and the purple dotted line represent the GOE $\langle r_{n} \rangle = 0.53$, GUE $\langle r_{n} \rangle = 0.60$, and GSE $\langle r_{n} \rangle = 0.67$ level statistics, respectively.}
    \label{fig:level_statistics_L_10}
\end{figure}
\begin{figure}
    \centering
    \includegraphics[width=\linewidth]{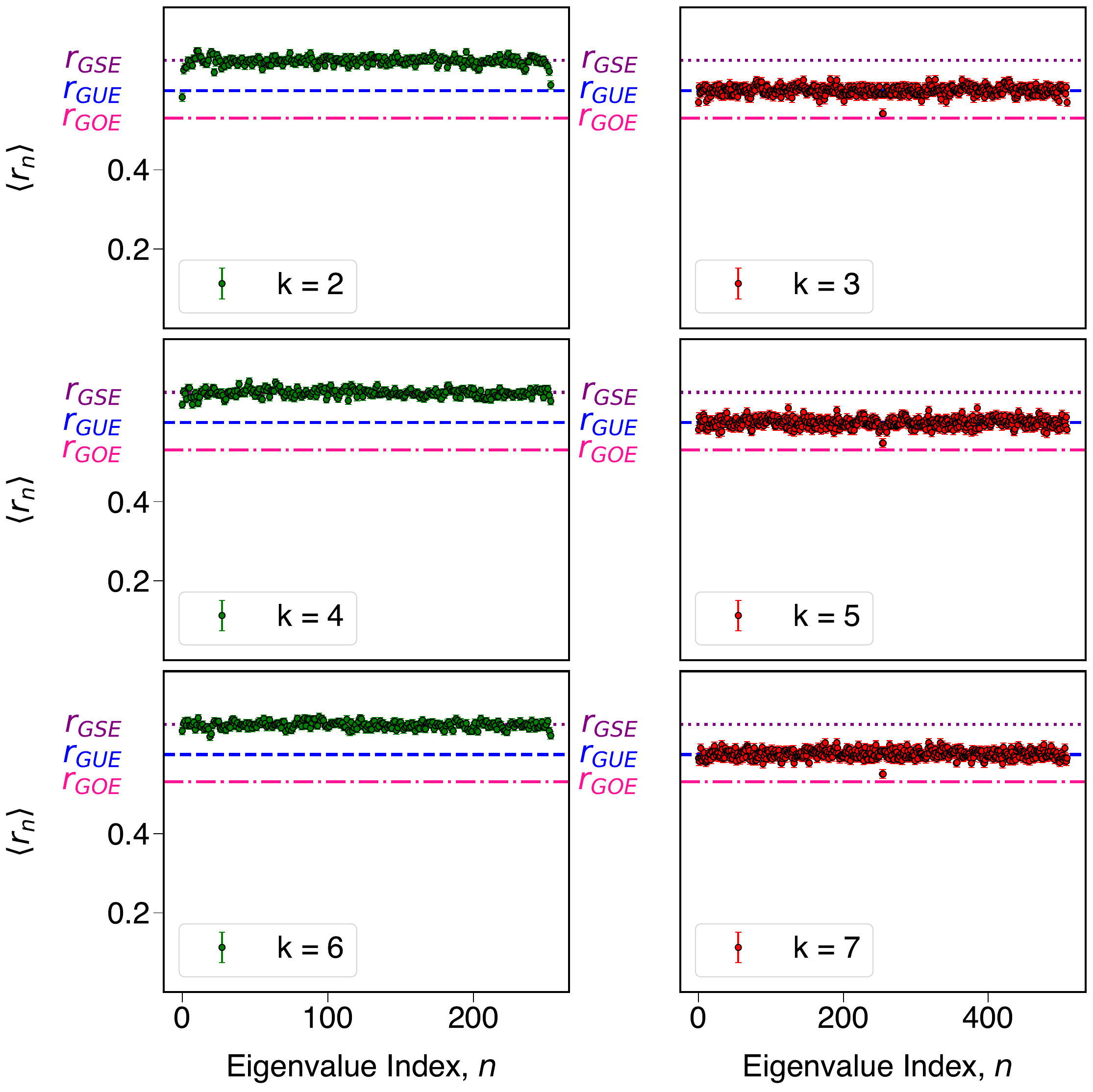}
    \caption{Average level statistics $\langle r_{n} \rangle$ of the Hamiltonian $H(0,1)$ for $L=9$ for $k=2,3,4,5,6,7$. The error bar indicates the standard error of the mean for $480$ different random realizations of the couplings. The pink dashed dot line, the blue dashed line, and the purple dotted line represent the GOE $\langle r_{n} \rangle = 0.53$, GUE $\langle r_{n} \rangle = 0.60$, and GSE $\langle r_{n} \rangle = 0.67$ level statistics, respectively. In the case of odd $L$ and even $k$, the eigenvalues of the Hamiltonian are doubly degenerate. The level statistics in that case are calculated by choosing one of the degeneracy sectors, \emph{i.e.}, the eigenvalues with either the odd or even indices.}
    \label{fig:level-statistics-L-9}
\end{figure}
\begin{figure}
    \centering
    \includegraphics[width=\linewidth]{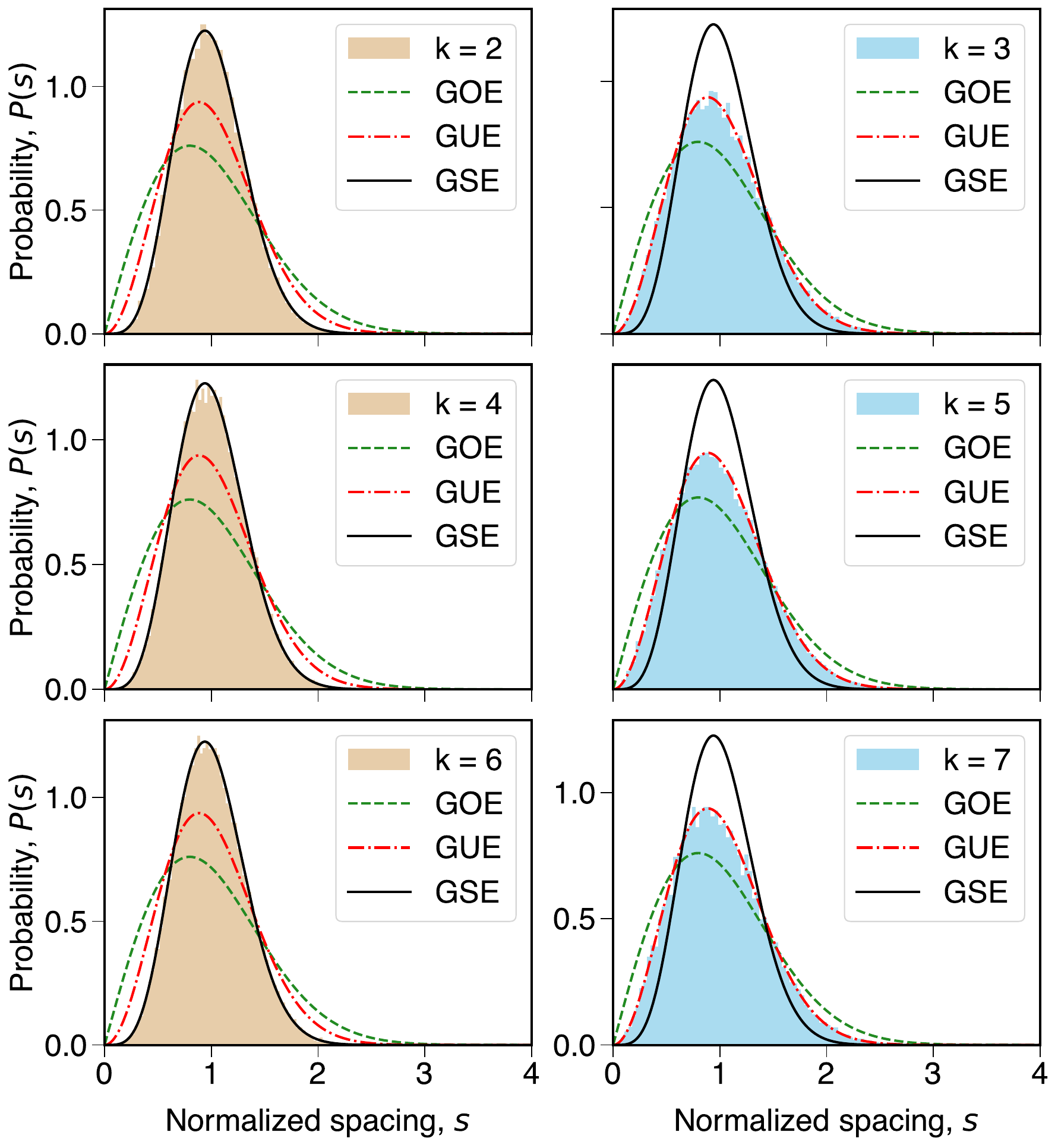}
    \caption{Average normalized spacing distribution of the Hamiltonians $H(0,1)$ for $L=9$ for locality $k=2,3,4,5,6,7$ averaged over $480$ random couplings.}
    \label{fig:normalized_energy_spacing_10}
\end{figure}
\begin{figure}
    \centering
    \includegraphics[width=\linewidth]{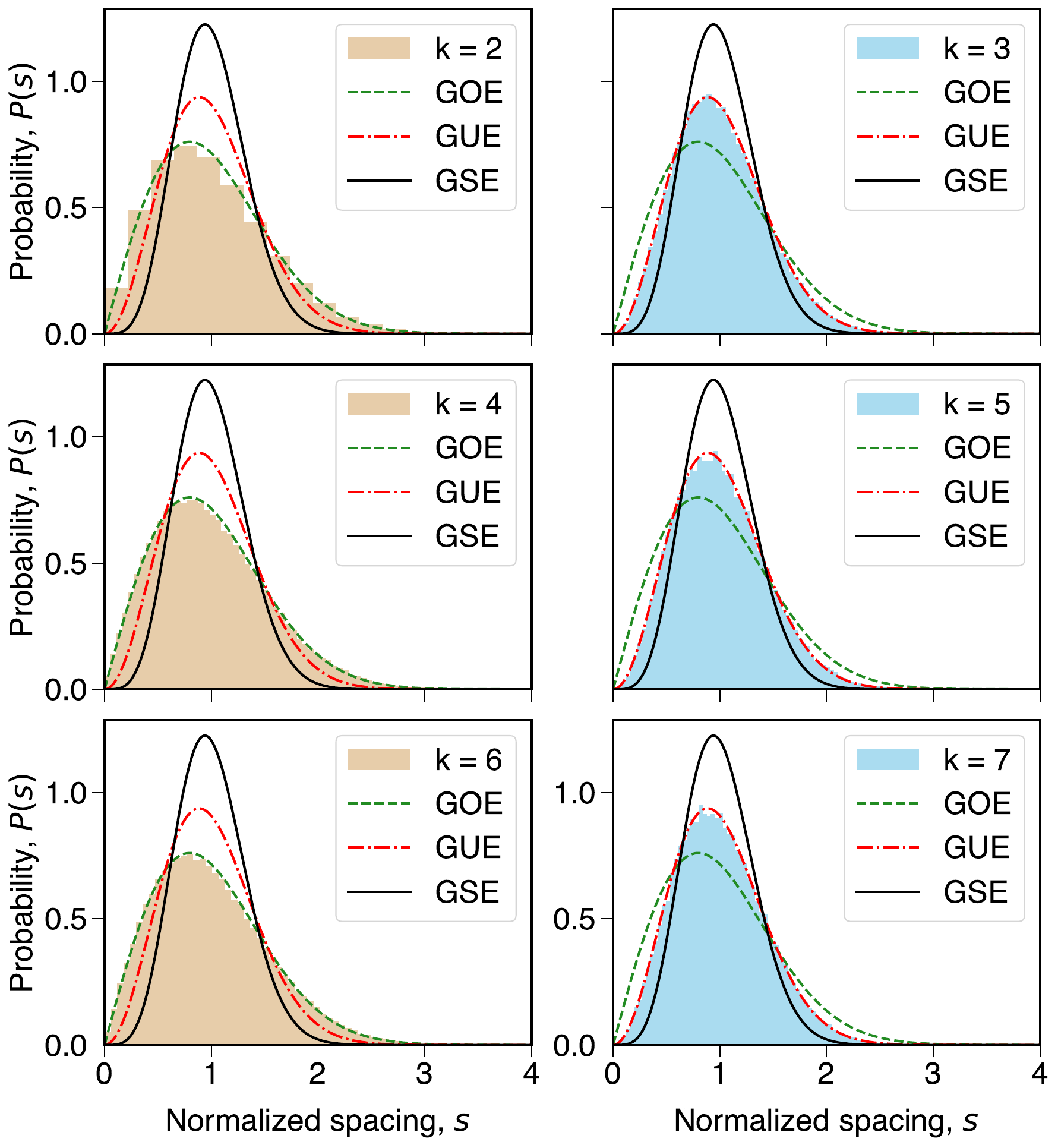}
    \caption{Average normalized spacing distribution of the Hamiltonians $H(0,1)$ for $L=10$ for locality $k=2,3,4,5,6,7$ averaged over $480$ random couplings. For even $k$, we choose one degeneracy sector to calculate the normalized spacings.}
    \label{fig:normalized_energy_spacing_9}
\end{figure}
We first present our numerical evidence for the dependence of the energy level statistics of the $k-$local Hamiltonians on the system size $L$ and locality $k$, as summarized in Table. \ref{tab:yourlabel}. In Figures \ref{fig:level_statistics_L_10} and \ref{fig:level-statistics-L-9}, we plot the energy level statistics of the Hamiltonian $H(0,1)$ averaged over random realizations of its couplings. For even $L$ and even $k$, the Hamiltonians belong to the GOE ensemble. For odd $L$ and even $k$, we found that the eigenvalues of $H(0,1)$ are doubly degenerate, suggesting the presence of Kramers degeneracy, and Hamiltonians belong to the GSE ensemble. When the locality $k$ is odd, the Hamiltonian belongs to the GUE ensemble for both odd and even system sizes. For odd $k$, we also found that the eigenvalue spectrum is symmetric around zero; for every eigenvalue $E$, there is an eigenvalue $-E$. This suggests that there is a unitary/anti-unitary operator $S$ that anticommutes with these Hamiltonians $H$, \emph{i.e.} $\{H, S \}= HS + SH = 0$. Since the energy level statistics of these Hamiltonians are robust to the coefficients and do not require any specific constraints on the structure in the Hamiltonian, the operator $S$ has to be defined in such a way that it anticommutes with $H(0,1)$ for any realization of its coefficients. In Figures \ref{fig:normalized_energy_spacing_10} and \ref{fig:normalized_energy_spacing_9}, we plot the probability distribution of the normalized spacings for different values of $L$ and $k$. We observe the same dependence on $L$ and $k$ in determining the random matrix ensembles of the Hamiltonians $H(0,1)$ as in Figure \ref{fig:level_statistics_L_10} and \ref{fig:level-statistics-L-9}.

\subsection{Time-reversal symmetry}
We now show that all the observed properties in energy level statistics of these Hamiltonians follow from whether it has time-reversal (TR) symmetry. A Hamiltonian $H$ has time-reversal symmetry if
\begin{eqnarray}
    T^{-1}HT=H,\;T=U_{T}K,\; U_{T}U^{*}_{T}=\pm \mathbf{1},\label{eq:time_reversal_definition}
\end{eqnarray}
where $U_{T}$ is a unitary operator, and $\textbf{1}$ is the identity matrix \cite{RevModPhys.88.035005}. We now construct the time-reversal operator explicitly for the $k-$local Hamiltonians. Under the action of the time-reversal operator, we expect the spin operator to change its sign $T^{-1}\Vec{S}T=-\Vec{S}$ or equivalently $T^{-1}\sigma_{j}T=-\sigma_{j}$, for $j=x,y,z$, since time-reversal reverses the angular momentum. For a single spin$-1/2$ particle, the time-reversal operator is given by $T=-i\sigma_{y}K$. Here, $K$ is the complex conjugation operator whose actions on the Pauli matrices are
\begin{align}
    K^{-1}\sigma_{x}K&=\sigma_{x}, K^{-1}\sigma_{y}K&=-\sigma_{y},K^{-1}\sigma_{z}K&=\sigma_{z}.\label{eq:conjugation_by_complex}
\end{align}
Generalizing this to the $k-$local Hamiltonian case, we define the time-reversal operator as
\begin{eqnarray}
    T &=& \bigotimes^{L}_{j=1} (-i\sigma^{(j)}_{y})K.\label{eq:time-reversal-definition}
\end{eqnarray}
The action of this operator on a $k-$local term is given by
\begin{eqnarray}
    &&T^{-1}(\sigma^{(j_{1})}_{\alpha_{1}}\otimes\sigma^{(j_{2})}_{\alpha_{2}}\otimes
    \cdots \otimes\sigma^{(j_{k})}_{\alpha_{k}})T\nonumber\\
    &&\hspace{0.5 cm}=(-1)^{k}\;(\sigma^{(j_{1})}_{\alpha_{1}}\otimes\sigma^{(j_{2})}_{\alpha_{2}}\otimes
    \cdots \otimes\sigma^{(j_{k})}_{\alpha_{k}}).
\end{eqnarray}
From this, we conclude that
\begin{eqnarray}
    T^{-1}HT &=& (-1)^{k}H. \label{eq:action-of-sigma_y-K}
\end{eqnarray}
This implies that $H$ is invariant under $T$ only if $k$ is even. For spin$-1/2$ particles, the time-reversal operator satisfies $T^{2}=-1$, which gives rise to Kramers degeneracy \cite{Sakurai_Napolitano_2020}. For the time-reversal operator defined in Eq.\eqref{eq:time-reversal-definition}, it satisfies
\begin{eqnarray}
    T^{2} &=& (-1)^{L}\mathbf{1}.\label{eq:sigma_y_K^2}
\end{eqnarray}
Thus, for odd $L$, $T^{2}=-\textbf{1}$, and for even $L$, $T^{2}=\textbf{1}$.

For odd values of $k$, time-reversal symmetry is absent, but the time-reversal operator satisfies $\{H, T\}=0$, from Eq.\eqref{eq:action-of-sigma_y-K}. From this, we conclude that the energy spectrum of $H$ for odd $k$ is symmetric around zero. With this, we are ready to explain all our observations discussed in section \ref{sec:results}.

\subsection{Classification of Hamiltonians}
We now explain the energy level statistics of the Hamiltonians defined in Eq.\eqref{eq:k_local_Hamiltonian_model} solely depending on the parity of $L$ and $k$. Below, we  explain each case separately.
\subsubsection{Odd L and even k}
For even $k$,  Eq.\eqref{eq:action-of-sigma_y-K} gives $[H, T]=0$. We now explicitly show that the eigenvalues of $H$ in this case are doubly degenerate. Suppose $|\psi\rangle$ be an eigenstate of $H$ with eigenvalue $\lambda$: $H|\psi\rangle=\lambda|\psi\rangle$. Applying $T$ from the left side on both sides of this equation gives: $TH|\psi\rangle=\lambda T|\psi\rangle$. Then using $TH=HT$, we get $H(T|\psi\rangle)=\lambda (T|\psi\rangle)$. Thus, the two states $|\psi\rangle$ and $T|\psi\rangle$ have the same eigenvalue if we can show $\langle \psi|T\psi\rangle=0$. We can proceed as follows: $\langle\psi|T\psi\rangle=\langle \psi TT|T\psi\rangle=\langle\psi T^{2}|T\psi\rangle=-\langle\psi|T\psi\rangle$. The first equality follows from the antilinearity of $T$ (\emph{i.e.} $\langle Ta|Tb\rangle=\langle b|a\rangle$), and the third equality follows because $L$ is odd. Thus, $\langle\psi|T\psi\rangle=0$. Together, $[H,T]=0$, and $T^{2}=-\textbf{1}$, imply that the Hamiltonians in this case belong to GSE.
\subsubsection{Even L even k}
Using the same time-reversal operator as for the previous case, we find: $[H,T]=0$, and $T^{2}=\textbf{1}$, implying that the Hamiltonian belongs to GOE.
\subsubsection{Even L odd k}
In this case, from Eq.\eqref{eq:action-of-sigma_y-K}, we find that $T$ does not commute with $H$ (rather, it anti-commutes with $H$). Therefore, the system does not have time-reversal symmetry, which implies that it belongs to the GUE. Since $k$ is odd, the energy spectrum of $H$ is symmetric around zero.
\subsubsection{Odd L odd k}
Similar to the previous case, odd $k$ implies that $\{T,H\}=0$. Thus, the energy spectrum is symmetric around zero. The absence of time-reversal symmetry implies that the Hamiltonians belong to GUE.\par
\textit{Remarks} -- For Hamiltonians whose matrix in the computational $\{|\uparrow\rangle, |\downarrow\rangle\}$ basis is purely real -- equivalently, built only from $\sigma_{0},\sigma_{x},\sigma_{z}$ or with an even number of $\sigma_{y}$ matrices in each term, the anti-unitary symmetry $T=K$ commutes with the Hamiltonian regardless of the value of $L$ and $k$. Therefore, these Hamiltonians belong to the GOE.

In addition to models with a single locality $k$, one can consider Hamiltonians composed of terms with different localities, say $k_{1}$ and $k_{2}$.  The symmetry classification in this case is governed by the combined effect of these terms. If either $k_{1}$ or $k_{2}$ is odd, the Hamiltonian lacks time-reversal symmetry and the level statistics is GUE. Conversely, if both localities are even, time-reversal symmetry is preserved and the statistics remain in the GOE or GSE depending on the parity of $L$. More generally, the random matrix ensemble is determined by whether time-reversal symmetry survives under the mixture of interaction terms.

Having characterized the $k-$local Hamiltonian with $\mu = 0$, we now shift our focus to the case where the couplings have a nonzero mean $(\mu \neq 0)$.  In the following section, we demonstrate that introducing a finite mean in the interaction strengths fundamentally alters the spectral properties, leading to the emergence of an energy gap between the ground state and the first excited state under specific conditions.

\subsection*{Part II: Energy gap and critical disorder scaling} \label{subsec:energy_gap}
In this section, we investigate the energy gap between the ground state and the first excited state for $k$-local Hamiltonians whose couplings have a nonzero mean, specifically for $\mu < 0$.
Rearranging Eq. \eqref{eq:H-mu-sigma-decomposition}, we obtain
\begin{eqnarray}
    H^{(L, k)}(\mu,\sigma^{2}) &=& \frac{L}{\sqrt{3^{k}} \binom{L}{k}} \bigg[\mu H^{(L, k)}(1,0) +\nonumber\\
    && \hspace{1 cm} \sigma \sqrt{\binom{L}{k}} H^{(L, k)}(0,1)\bigg], \label{eq:normalized_single_parameter_hamiltonian}
\end{eqnarray}
where we have introduced superscripts to indicate the system size and locality explicitly. In the following, we keep $\mu$ fixed, thereby reducing the Hamiltonian to a one-parameter family characterized solely by $\sigma$, which we refer to as the \textit{disorder strength}. We then examine how this disorder effects the energy spectrum. As a starting point, we analyze the spectrum and eigenstates of the disorder-free Hamiltonian ($\sigma = 0$).

\subsection{Eigenstructure in the absence of disorder} \label{subsec:eigenstructure_in_the_absence_of_disorder}
The  deterministic or the `disorder-free' part of the Hamiltonian $H^{(L, k)}(\mu, 0)$ has the eigenvalues 
\begin{eqnarray}
    \lambda^{(L, k)}_{n} &=& \mu L \binom{L}{k}^{-1} \sum^{k}_{j = 0} (-1)^{j} \binom{n}{j} \binom{L - n}{k - j} ,\\
    &=& \mu L \binom{L}{k}^{-1} K^{L, 2}_{k}(n),\label{eq:analytical_eigenvalues_of_disorder_free_Hamiltonian}
\end{eqnarray}
for $n = 0, 1, 2, \cdots, L$, and each eigenvalue $\lambda^{(L, k)}_{n}$ has a degeneracy of $\binom{L}{n}$. The functions $K^{L, 2}_{k}(n)$ are called Krawtchouck polynomials \cite{Huffman_Pless_2003, kravchuk1929}. The eigenstates corresponding to these eigenvalues are the product states
\begin{eqnarray}
    |\Psi_{s}\rangle &\equiv & \bigotimes^{L}_{j = 1} |u_{s_{j}} \rangle ,\label{eq:exact_eigenstates_of_H_1_0}
\end{eqnarray}
where $s = (s_{1}, s_{2}, \cdots, s_{L})$ with $s_{j} \in \{ +, -\}$, $|u_{\pm}\rangle$ are the eigenstates of $(\sigma_{x} + \sigma_{y} + \sigma_{z})$ given by
\begin{eqnarray}
    |u_{+} \rangle &=& \frac{1}{\sqrt{2}}\bigg(  \sqrt{1 - \frac{1}{\sqrt{3}}} |0\rangle - e^{i\pi/4}\sqrt{1 + \frac{1}{\sqrt{3}}}|1\rangle\bigg),\quad \\
    |u_{-} \rangle &=& \frac{1}{\sqrt{2}}\bigg(  \sqrt{1 + \frac{1}{\sqrt{3}}} |0\rangle + e^{i\pi/4}\sqrt{1 - \frac{1}{\sqrt{3}}}|1\rangle\bigg) .\quad  \label{eq:eigenstates_of_sigma_x+sigma_y+sigma_z}  
\end{eqnarray}
The index $n$ in the eigenvalues in Eq.\eqref{eq:analytical_eigenvalues_of_disorder_free_Hamiltonian} corresponds to the number of $s_{j} = -1$ in Eq. \eqref{eq:exact_eigenstates_of_H_1_0}. The Krawtchouck polynomials satisfy $K^{L, 2}_{k} (n) = (-1)^{k} K^{L, 2}_{k} (n)$, which implies that for even $k$ the extremum (maximum if $\mu > 0$ and minimum if $\mu < 0$) eigenvalue occurs at $n=0, L$. This corresponds to two degenerate eigenstates: $s=(+, +, \cdots, +)$ or $n = 0$, and $s=(-, -, \cdots, -)$ or $n = L$. When $k$ is odd, the spectrum is symmetric around zero, and the maximum and the minimum occur at $n = L$ and $n = 0$ respectively for $\mu < 0$. For a given system size $L$, all the disorder-free $k-$local Hamiltonians have the same eigenstates given by Eq.\eqref{eq:exact_eigenstates_of_H_1_0}, which follows from the property: $[H^{(L,k_{1})}(1, 0), H^{(L,k_{2})}(1, 0)] = 0$, for any $1 \leq k_{1}, k_{2} \leq L$. Eigenvalues of these systems have been previously studied in the context of the Hamming graph/scheme  \cite{vanLint_Wilson_2001}. In the next section, we examine the energy spectrum of these Hamiltonians for $\sigma \neq 0$.

\subsection{Spectrum of the disordered Hamiltonian}
\begin{figure}
    \includegraphics[width=\linewidth]{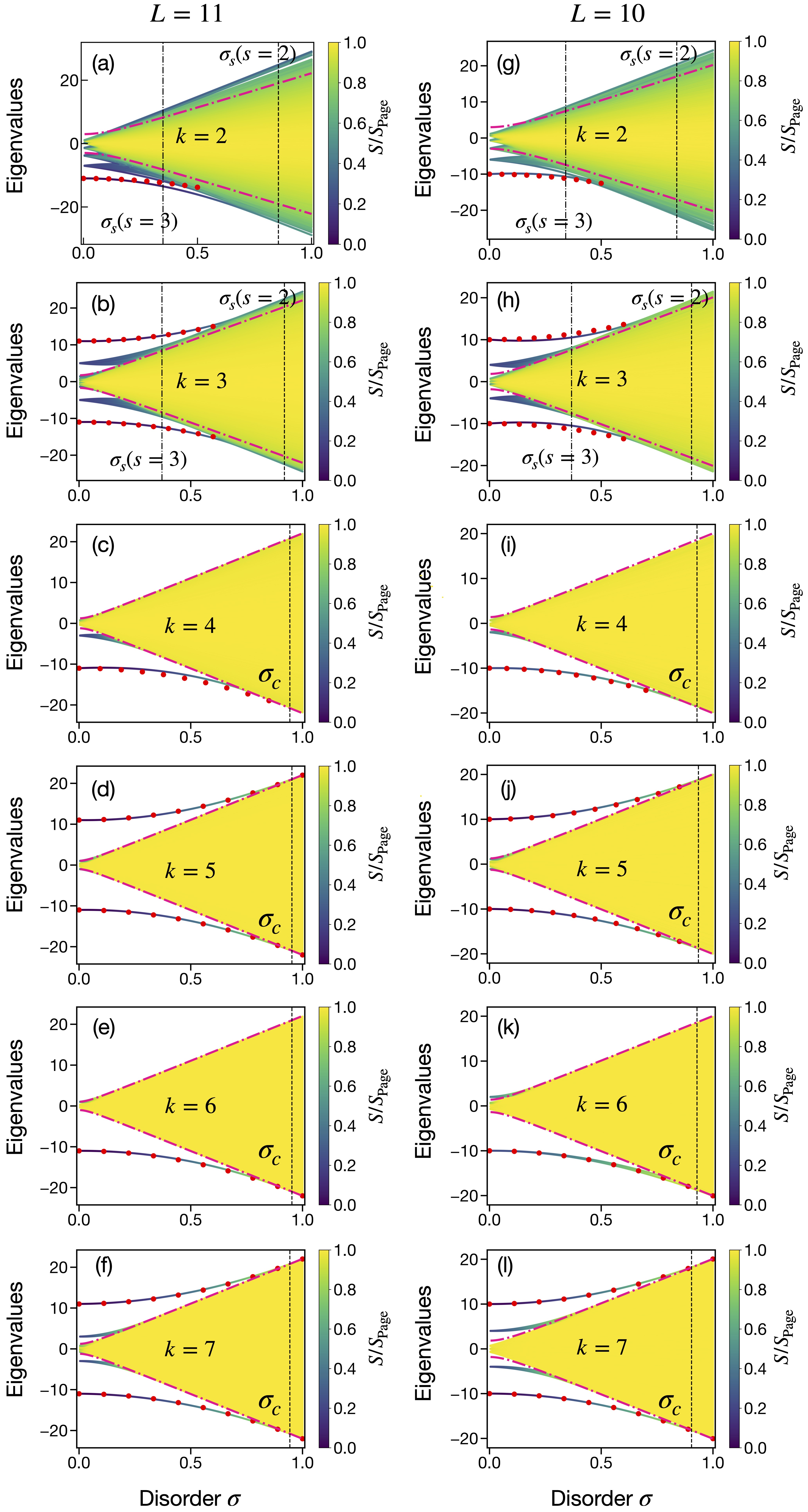}
\caption{\label{fig:spectrum_L_10_9} Plot of the energy spectra of the Hamiltonian defined in Eq.~\eqref{eq:normalized_single_parameter_hamiltonian} as a function of disorder strength $\sigma$ for $\mu = -1$, shown for system sizes $L = 11$ and $L = 10$, each corresponding to a single realization of random couplings. Panels (a)–(f) correspond to $L = 11$ with $k=2, 3, 4, 5, 6, 7$, respectively; panels (g)–(l) show the same for $L=10$. The color scale represents the half-system von Neumann entanglement entropy normalized by its Page value \cite{PhysRevLett.71.1291}: $S/S_{\mathrm{Page}}$, where $S_{\mathrm{Page}} \approx 3.22$ for $L=11$ (using a bipartition into $L_A = 6$, $L_B = 5$) and $S_{\mathrm{Page}} \approx 2.97$ for $L=10$. The two dashed-dotted pink lines in each plot represent the envelope given by Eq.\eqref{eq:radius_of_wigner_semicircle_as_function_of_sigma}. The red dotted curves represent Eq. \eqref{eq:extremum_eigenvalue} for the ground state energy for even $k$ and the ground state energy and the highest energy for odd $k$. For $k = 2, 3$, we have shown the critical disorder values predicted by Eq.\eqref{eq:sigma_c_for_a_general_s} using vertical black lines corresponding to the ground state merging into the bulk that are within two $(s = 2)$ and three $(s = 3)$ standard deviations from the mean. For $k > 3$, the critical disorder predicted by Eq.\eqref{eq:critical_disorder_approximate_formula_k_L/2} is shown with a vertical dotted line.}
\end{figure}

We now discuss the energy spectrum of $H^{(L,k)}(\mu, \sigma^{2})$ for nonzero disorder $\sigma$, keeping the mean $\mu$ fixed. Since our focus is on the energy gap between the ground state and the first excited state, we set $\mu < 0$ in the following. In Figure \ref{fig:spectrum_L_10_9}, we plot the energy spectrum/eigenvalues of $H^{(L, k)}( \mu, \sigma^{2})$ as a function of $\sigma$ for $L = 10, 11$ with $k = 2, 3, 4, 5, 6, 7$ -- obtained numerically using exact diagonalization for a single disorder realization with $\mu = -1$. For clarity, we refer to the eigenvalues in the middle of the spectrum as the \textit{bulk}, and the ones that lie far at the extreme edges of the spectrum simply as the \textit{ground state} or the \textit{highest excited state}.
The degeneracies of the bulk eigenvalues are lifted in the presence of nonzero disorder $\sigma$, and the edges of the bulk spread proportionally to the disorder strength. We show later that this spreading is linear in $\sigma$ for $k \gg \sqrt{L}$. In the same plot, we indicate the bipartite von Neumann entanglement entropy of each eigenstate through colorbars, which diagnoses thermalization in the bulk and the maximum/minimum eigenstates. For even $L$, we choose a subsystem A of $L/2$ consecutive sites: $L_{A} = L_{B} = L /2$, for odd $L$, we choose the subsystems as: $L_{A} = \lfloor L /2 \rfloor, L_{B} = L - L_{A}$ with periodic boundary conditions. We then calculate the entropy as $S_{A} = - \mathrm{Tr}(\rho_{A} \ln \rho_{A})$, where $\rho_{A} = \mathrm{Tr}_{\mathrm{\Bar{A}}}(|\psi\rangle \langle \psi|)$ is the reduced density matrix of eigenstate $|\psi\rangle$. We then average over $\lfloor L/2 \rfloor$ independent choices of the subsystem $A$ for all given eigenstates.

\begin{figure}
    \centering
    \includegraphics[width=\linewidth]{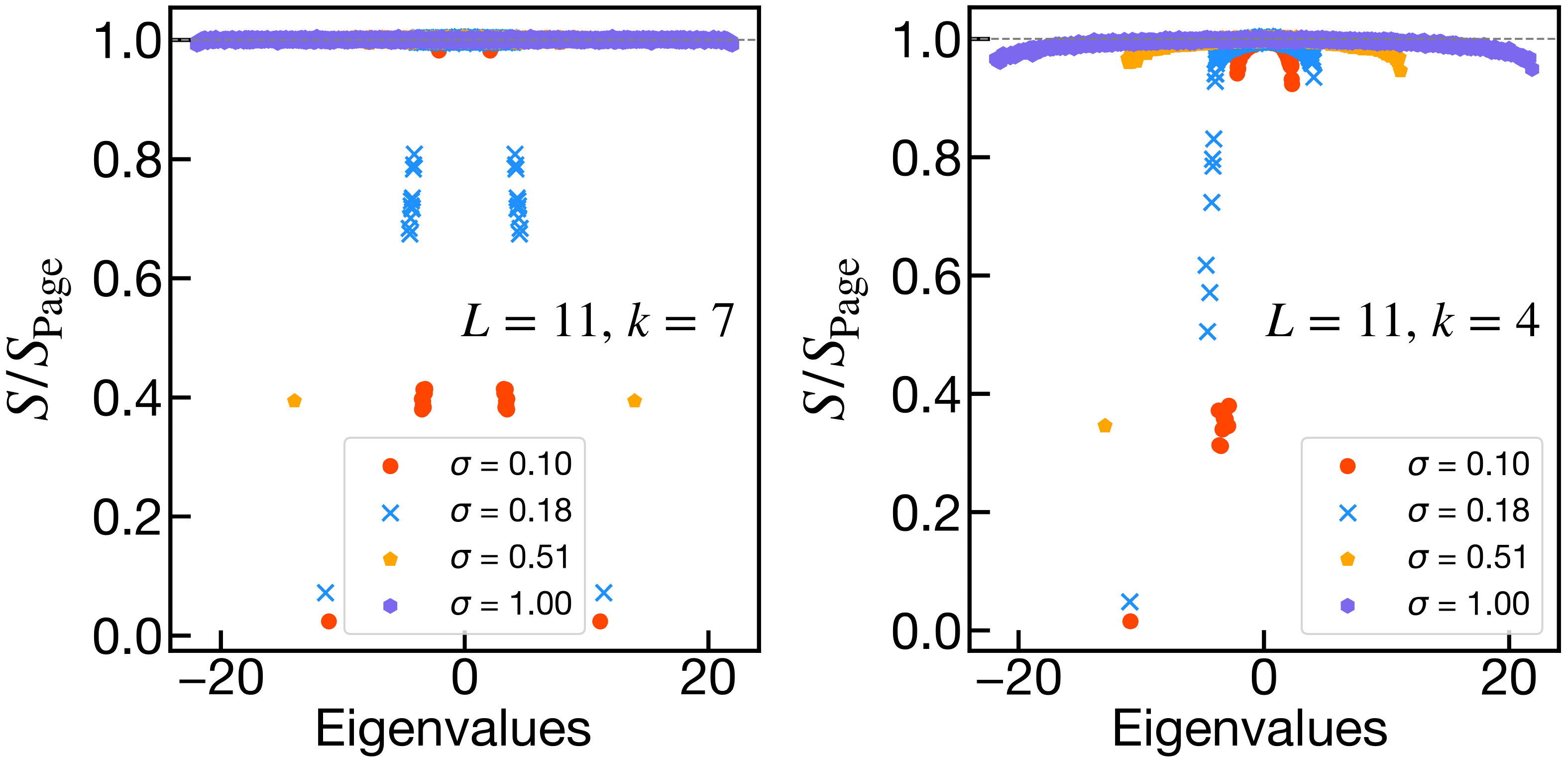}
    \caption{Half-system entanglement entropy normalized by the Page value, plotted against the eigenvalues of $H^{(L, k)}(\mu, \sigma^{2})$ with $L = 11, k = 4, 7$ for different disorder strengths $\sigma$.}
    \label{fig:entanglement_entropy_vs_eigenvalues_different_sigma_values}
\end{figure}

We find that the bulk thermalizes quickly at small disorder strength and reaches the maximum entanglement entropy given by the Page value $S_{\mathrm{Page}} = \lfloor L \ln2 - 1 \rfloor$ \cite{PhysRevLett.71.1291}. The maximum/minimum eigenstates that lie far from the bulk display low entropy, retaining memory of their unperturbed states until a critical value of disorder strength, say $\sigma_{c}$, at which they merge in the bulk. Since the density of states in general vanishes near the spectral edges, the eigenstate thermalization hypothesis (ETH) is not expected to hold in this regime. Consequently, the observation that eigenstates at the spectral edges are nonthermal is consistent with theoretical expectations \cite{PhysRevE.90.052105}. To make the entanglement structure more explicit, we plotted the normalized entropy as a function of the eigenvalues in Figure \ref{fig:entanglement_entropy_vs_eigenvalues_different_sigma_values} for $L =11$ and $k = 4, 7$ for several values of $\sigma$.

From Eq. \eqref{eq:normalized_single_parameter_hamiltonian}, we expect that for small $\sigma$, the deterministic component $H^{(L,k)}(\mu, 0)$ dominates the disordered term $H^{(L,k)}(0, \sigma^{2})$. Consequently, the energy level statistics at small $\sigma$ are expected to deviate from the canonical random matrix ensembles such as GOE, GUE, or GSE. For large $\sigma$, the disordered component dominates, and we expect random matrix statistics from $H^{(L,k)}(\mu, \sigma^{2})$. Latter in this section, we establish that random matrix statistics are expected from this model at exponentially small disorder strength $\sigma \sim 2^{-L}$.

The density of states of $H^{(L, k)}(0, 1)$ changes from a Gaussian distribution for $k \ll \sqrt{L}$, to the Wigner semicircular distribution when $k \gg \sqrt{L}$ \cite{Erdős2014, FRENCH19715, RevModPhys.53.385}. For sufficiently large $\sigma$, we expect the density of states of $H^{(L, k)}(\mu, \sigma^{2})$ to follow the same crossover behavior. The Wigner semicircle law is given by \cite{c9eb8278-b5e2-37bc-a322-8e81785f98ed, Potters_Bouchaud_2020}
\begin{eqnarray}
       \rho_{\mathrm{sc}}(E) = \begin{cases} \frac{2}{\pi R^{2}} \sqrt{ R^{2} - E^{2}} , & \text{for } \quad |E| < R  \\
       0, & \text{otherwise} \label{eq:Wigner-semicircle_law_H_1_0}
\end{cases}
\end{eqnarray}
where $R = 2\;(\mathrm{standard\; deviation\; of\; eigenvalues})$. To determine $R$ as a function of disorder $\sigma$, we calculate the variance of the eigenvalues of $H^{(L,k)}(\mu, \sigma^{2})$, with $\langle \cdot \rangle$ denoting average over the random coefficients
\begin{eqnarray}
    \mathrm{Var} &\equiv & \frac{1}{2^{L}} 
    \langle \mathrm{Tr}[(H^{(L, k)}(\mu, \sigma^{2}))^{2}] \rangle , \\
    &=& \frac{1}{2^{L}} \bigg\langle \sum_{a, b} J_{a} J_{b} \mathrm{Tr}[P_{a} P_{b}] \bigg\rangle ,\\
    &=& \sum_{a} \langle J^{2}_{a} \rangle ,\\
    &=& 3^{k} \binom{L}{k} ([E(J_{a})]^{2} + \mathrm{Var}(J_{a})) ,\\
    &=& 3^{k} \binom{L}{k} \bigg( \frac{\mu^{2} L^{2}}{3^{k} \binom{L}{k}^{2}} + \frac{\sigma^{2} L^{2}}{3^{k} \binom{L}{k}}\bigg), \\
    &=& L^{2} \bigg(\sigma^{2} + \mu^{2} \binom{L}{k}^{-1}\bigg),
\end{eqnarray}
where we have used $\mathrm{Tr}[P_{a}P_{b}] = 2^{L} \delta_{a, b}$, for two $k-$local Pauli terms $P_{a}, P_{b}$. From this, we obtain the radius of the Wigner semicircle in Eq. \eqref{eq:Wigner-semicircle_law_H_1_0} as
\begin{eqnarray}
    R (\sigma) &=& 2 L \sqrt{\sigma^{2} + \mu^{2} \binom{L}{k}^{-1}} .\label{eq:radius_of_wigner_semicircle_as_function_of_sigma}
\end{eqnarray}
From this equation, we see that the energy spectrum of $H^{(L,k)}(\mu, \sigma^{2})$ approximately grows linearly in $\sigma$. The two functions, $R(\sigma)$ and $- R(\sigma)$, envelop the bulk eigenvalues of the Hamiltonian $H^{(L,k)}(\mu, \sigma^{2})$ for $k \gg \sqrt{L}$ as shown in Figure \ref{fig:spectrum_L_10_9}.

To determine the eigenvalues lying outside the Wigner semicircle envelope, \textit{i.e.} the ground state and the highest excited state for $\sigma \neq 0$, we start by writing
 Eq.\eqref{eq:normalized_single_parameter_hamiltonian} in the eigenbasis of $H^{(L, k)}(\mu, 0)$:
\begin{eqnarray}
    H^{(L,k)}(\mu, \sigma^{2}) &\equiv & D + (\sigma L) V, \label{eq:H_in_eigenbasis_of_H_1_0}
\end{eqnarray}
    where $D = \mathrm{diag}(\lambda^{(L,k)}_{0}, \lambda^{(L,k)}_{1}, \cdots, \lambda^{(L,k)}_{2^{L}-1})$ with $\lambda^{(L,k)}_{n}$ given by Eq. \eqref{eq:analytical_eigenvalues_of_disorder_free_Hamiltonian}, $V \equiv U^{\dagger} H^{(L,k)}(0,1) U / \sqrt{3^{k} \binom{L}{k}}$, and $U$ is the matrix that diagonalizes $H^{(L,k)}(\mu, 0)$. When $k$ is odd, then $U$ is unitary, so $V$ is a GUE. Similarly, when $k$ is even and $L$ is even (odd), it can be shown that $H^{(L, k)}(\mu, 0)$ can be made real using an orthogonal transformation \cite{Haake2018QuantumSignatures}, therefore $V$ is GOE/GSE. Thus, in this new basis, the random matrix ensemble associated with $H^{(L, k)}(0, 1)$ remains unchanged.

    Eq.\eqref{eq:H_in_eigenbasis_of_H_1_0} can also be interpreted as a case of the Rosenzweig–Porter model \cite{PhysRev.120.1698, Čadež_2024} which has the form $H = H_{0} + V/2^{L\gamma/2}$, where $H_{0}$ is a diagonal matrix with elements chosen from a Gaussian distribution, and $V$ is a random matrix. The RP model shows Wigner-Dyson statistics for $\gamma < 1$, fractal phases when $1 < \gamma < 2$, and localized statistics for $\gamma > 2$ \cite{PhysRevB.108.L060203, PhysRevLett.76.2258, 10.1063/1.531918, Kravtsov2015, Pino_2019, PhysRevE.56.1471}. Drawing an analogy between the RP and the model in Eq.\eqref{eq:H_in_eigenbasis_of_H_1_0}, we find that our model should show random matrix statistics for exponentially small disorder $\sigma \sim 2^{ - L}$ for $\mu \sim \mathcal{O}(1)$.
    
\begin{figure}[h]
    \centering
    \includegraphics[width=\linewidth]{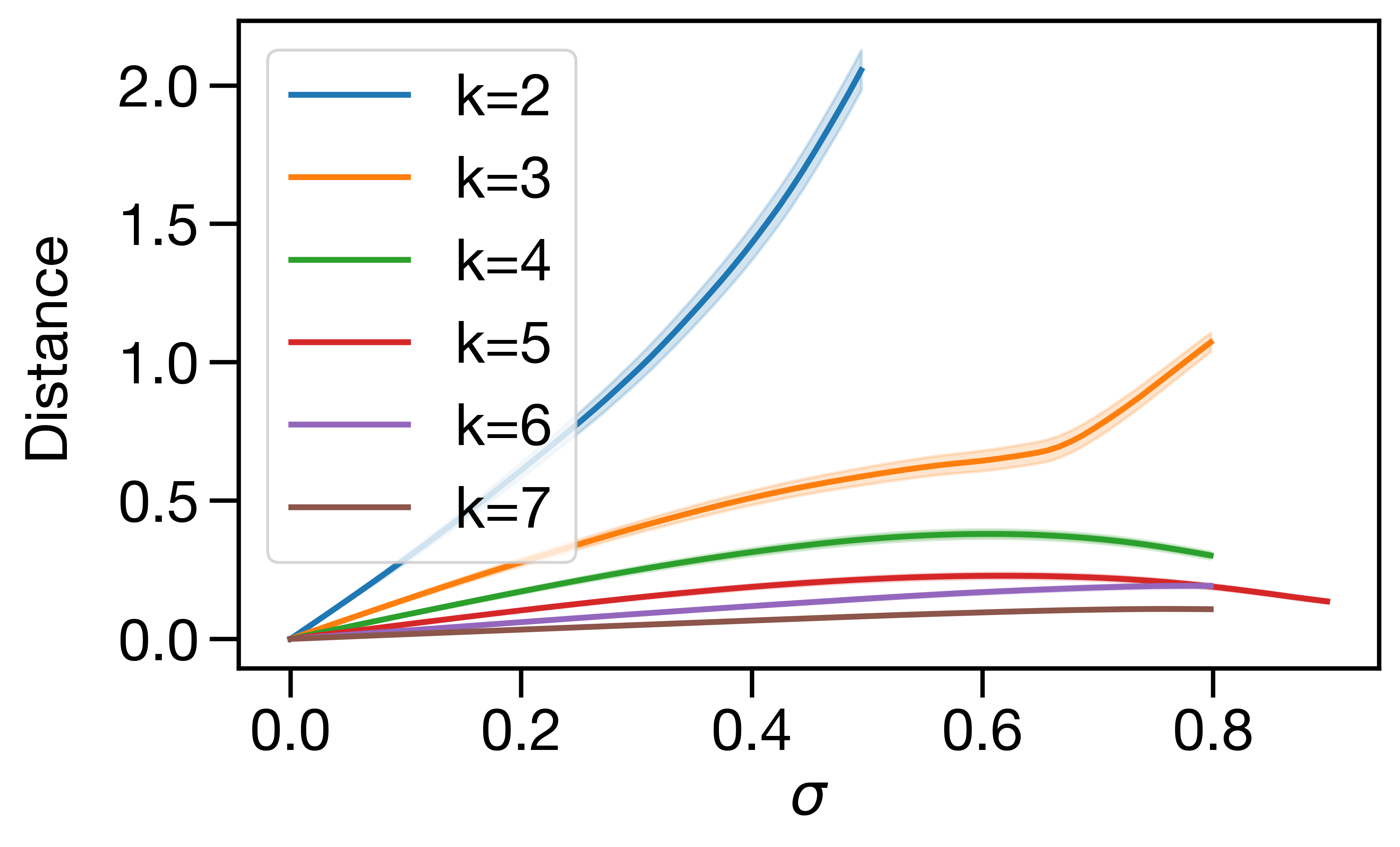}
    \caption{Plot of the absolute difference/ distance between the maximum eigenvalues calculated using Eq. \eqref{eq:extremum_eigenvalue} and using exact diagonalization, as a function of disorder strength $\sigma$, for $L=10, k = 2, 3, 4, 5, 6, 7$. Error bars / bands indicate the standard error of the mean over $256$ different realizations of the coefficients of the Hamiltonian. For each $k$ the difference is calculated up to the point the ground state remains detached from the bulk.}
    \label{fig:distance_analytical_calculated_in_python}
\end{figure}

    Matrices of the form defined by Eq. \eqref{eq:H_in_eigenbasis_of_H_1_0} are sometimes referred to as \emph{deformed random matrices} \cite{capitaine2016spectrumdeformedrandommatrices}, and various results exist for the distribution of the largest eigenvalue of such deformed matrices when the matrix $D$, commonly referred to as the perturbation, has different ranks \cite{SFEdwards_1976, 10.1214/009117905000000233, Péché2006, 10.1088/1751-8121/adf835, BENAYCHGEORGES2011494, 10.1214/08-AOP394, capitaine2011centrallimittheoremseigenvalues, capitaine2011freeconvolutionsemicirculardistribution, Féral2007, pizzo2011finiterankdeformationswigner, renfrew2012finiterankdeformationswigner}. For instance, when the perturbation matrix $D$ has rank-one, the spectrum exhibits the well-known Baik–Ben Arous–Péché (BBP) transition -- as the strength of the perturbation is increased, the largest eigenvalue detaches from the bulk and becomes an outlier \cite{Péché2006}. For our case, we use the result for a full rank perturbation \cite{10.1088/1751-8121/adf835}: for $\mu < 0$ we can write, on average, the ground state energy of $H^{(L, k)}(\mu, \sigma^{2})$ as
\begin{eqnarray}
    \lambda_{\mathrm{min}}(\sigma) &=& \mu L + \sigma^{2} L^{2} \int \frac{\rho(v)}{\mu L   - v} dv ,\label{eq:maximum_eigenvalues_integral_formula}
\end{eqnarray}
where the density of states of $H^{(L,k)}(\mu, 0)$ is defined as
\begin{eqnarray}
    \rho(v) &=& \frac{1}{2^{L}} \sum_{j} \delta(v - v_{j}) .\label{eq:density_of_states_summation_definition}
\end{eqnarray}
Combining the above two equations, we obtain
    \begin{eqnarray}
        \lambda_{\mathrm{min}} (\sigma) &=& \mu L + \frac{\sigma^{2} L^{2}}{2^{L}} \sum_{v \neq \mu L} \frac{1}{\mu L - v} \;, \label{eq:extremum_eigenvalue}
    \end{eqnarray}
where $v$ runs over all the eigenvalues of $H^{(L,k)}(\mu, 0)$ except $\mu L$. For odd $k$, by symmetry we conclude $\lambda_{\mathrm{min}}(\sigma) = - \lambda_{\mathrm{max}}(\sigma)$. When $L$ and $k$ are both even, then for $\sigma = 0$ the ground state is doubly degenerate with eigenstates $|\psi_{n = 0} \rangle = |u_{-}\rangle^{\otimes L}$, and $|\psi_{n = L} \rangle = |u_{+} \rangle^{\otimes L}$. Although introducing disorder lifts the degeneracy between these two states, we prove in the appendix (\ref{sec:degeneracy_of_largest_eigenvalue_even_L_even_k}) that they are degenerate in the thermodynamic limit ($ L \to \infty)$. Hence, there is a unique ground state energy in each case that is given by Eq. \eqref{eq:extremum_eigenvalue} for all $L$ and $k$. In Figure \ref{fig:spectrum_L_10_9}, where we plot Eq. \eqref{eq:extremum_eigenvalue} on top of the spectra for different values of $L$ and $k$, which shows good agreement between them. In addition, Figure \ref{fig:distance_analytical_calculated_in_python} shows the absolute distance between the maximum eigenvalue calculated using Eq. \eqref{eq:extremum_eigenvalue} and the ones calculated using exact diagonalization for $L=10$ and $k=2, 3, 4, 5, 6, 7$. These two results confirm the validity of Eq. \eqref{eq:extremum_eigenvalue}. In the next section, we determine the critical disorder strength $\sigma_{c}$ at which the ground state merges with the bulk, thereby closing the energy gap.

\subsection{Critical disorder and its scaling}
As mentioned earlier, for $k \gg \sqrt{L}$, the bulk eigenvalues follow the Wigner semicircle law, resulting in a sharp cutoff at $- R$ and $R$ \footnote{For finite $L$, the largest eigenvalues do not terminate abruptly at this edge; instead, they exhibit fluctuations governed by Tracy–Widom statistics around a  region that goes as $N^{-2/3}$, where $N = 2^{L}$ \cite{Potters_Bouchaud_2020}. These fluctuations vanish in the thermodynamic limit}. In contrast,  for $k \ll \sqrt{L}$, the bulk follows a Gaussian distribution and lacks a sharp cutoff. Therefore, we divide our analysis into two regimes.

\emph{$k \gg \sqrt{L}$ limit} -- From Eq.\eqref{eq:radius_of_wigner_semicircle_as_function_of_sigma}, at disorder strength $\sigma$, the bulk eigenvalues lie within $R(\sigma) =\pm 2 L \sqrt{\sigma^{2} + \mu^{2} \binom{L}{k}^{-1}}$. Let $\sigma_{c}$ be the disorder value at which the ground state enters the bulk, then using Eq. \eqref{eq:radius_of_wigner_semicircle_as_function_of_sigma} and \eqref{eq:extremum_eigenvalue}, we can write
\begin{eqnarray}
    2 L \sqrt{\sigma^{2}_{c} + \mu^{2}/ \binom{L}{k}} &=& \mu L \\
    && \hspace{1 mm} + \frac{\sigma^{2}_{c} L^{2}}{2^{L}} \sum_{v \neq \mu L } \frac{1}{\mu L- v}, \label{eq:critical_disorder_exact_formula}
\end{eqnarray}
While this equation can be solved numerically for a given $L$ and $k$, here we are interested in the scaling of the critical disorder $\sigma_{c}$ with the system size $L$. Let us focus on the case of $k=L/2$ when $L$ is even and $k=(L - 1)/2$ when $L$ is odd. This choice of $k$ maximizes the binomial coefficient. When $k$ is odd, the energy spectrum has the maximum and the minimum eigenvalue at $-\mu L$ and $ \mu L$ respectively. The rest of the eigenvalues are concentrated around zero in the limit $L \to \infty$. When $k$ is even, the ground state energy (minimum eigenvalue) is at $\mu L$ with the rest concentrated around zero. Thus, in the limit of large $L$, we can approximate Eq. \eqref{eq:critical_disorder_exact_formula} as
\begin{eqnarray}
    2 \sqrt{\sigma^{2}_{c} + \mu^{2}/ \binom{L}{L/2}} &\approx & \mu + \frac{\sigma^{2}_{c}}{\mu}. \label{eq:critical_disorder_approximate_formula_k_L/2}
\end{eqnarray}
This gives the critical disorder strength for gap closing as
\begin{eqnarray}
    \sigma_{c}  \approx | \mu| \sqrt{1 - 2 \binom{L}{L/ 2}^{-1}} .
\end{eqnarray}
In the thermodynamic limit, this yields a constant critical disorder $\sigma_{c} = |\mu|$.

\vspace{0.5 cm}

\begin{figure}[h]
    \centering
    \includegraphics[width=\linewidth]{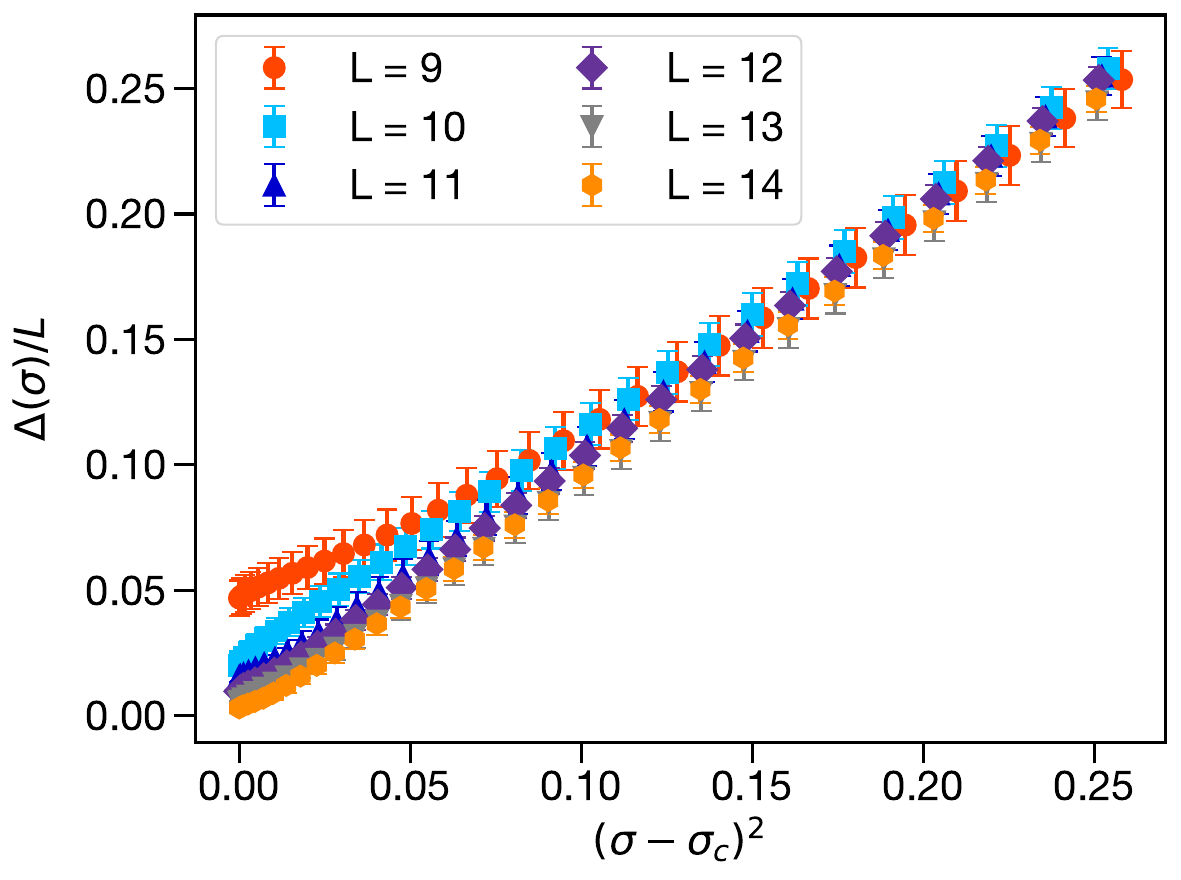}
\caption{Scaling of the bulk–edge energy gap near the critical point. The normalized gap $\Delta(\sigma)/L$ is plotted against  $(\sigma_{c} - \sigma)^{2}$  for system sizes $L = 9, 10, 11, 12, 13, 14$ at fixed $k=5$. The data collapse onto a common straight line, demonstrating the universal quadratic scaling $\Delta(\sigma)/ L \propto (\sigma_{c} - \sigma)^{\gamma}$ with critical exponent $\gamma = 2$. Error bars denote the standard error of the mean over ten disorder realizations.}
    \label{fig:edge_bulk_energy_gap_vs_sigma_for_L_10}
\end{figure}

\textit{Energy gap scaling in $k \gg \sqrt{L}$ limit} -- For $k \gg \sqrt{L}$, it is straightforward to show that the energy gap between the ground state and the first excited state (bulk) scales as $\Delta(\sigma) \approx L (\mu - \sigma)^{2} / |\mu|$. In Figure \ref{fig:edge_bulk_energy_gap_vs_sigma_for_L_10}, we plot the normalized energy gap $\Delta(\sigma)/L$ as a function of $(\sigma_{c} - \sigma)^{2}$ for $k=5$ and $L = 9, 10, 11, 12, 13, 14$. The data collapse onto a common straight line, demonstrating the universal quadratic scaling $\Delta(\sigma)/ L \propto (\sigma_{c} - \sigma)^{\gamma}$ with critical exponent $\gamma = 2$.


\vspace{0.5 cm}

\emph{$k \ll \sqrt{L}$ limit} -- 
Without a sharp cutoff of the bulk eigenvalues, it is difficult to obtain an exact value of the disorder at which the ground state enters the bulk and the energy gap closes. Instead, we define a disorder strength, denoted by $\sigma_{s}$, when the ground state enters into the bulk eigenvalues that lie within $s$ standard deviations from their mean. Using Eq.\eqref{eq:radius_of_wigner_semicircle_as_function_of_sigma} and \eqref{eq:extremum_eigenvalue}, we  can then write
\begin{eqnarray}
    s L \sqrt{\sigma^{2}_{s} + \mu^{2}/ \binom{L}{k}} &=& \mu L \\
    && \hspace{1 mm} + \frac{\sigma^{2}_{s} L^{2}}{2^{L}} \sum_{v \neq \mu L } \frac{1}{\mu L- v} \;. \label{eq:sigma_c_for_a_general_s}
\end{eqnarray}
Let us focus on a specific value of locality, say $k=2$, in the limit of large system size $L$. In the appendix (Sec. \ref{sec:critical_disorder_scaling_k_2_derivation}), we prove that irrespective of $s=2, (95.45 \%)$ or $s=3, (99.73 \%)$, Eq.\eqref{eq:sigma_c_for_a_general_s} gives the disorder scaling
\begin{eqnarray}
    \sigma_{s} \sim |\mu|.
\end{eqnarray}
\begin{figure}
    \centering
    \includegraphics[width=\linewidth]{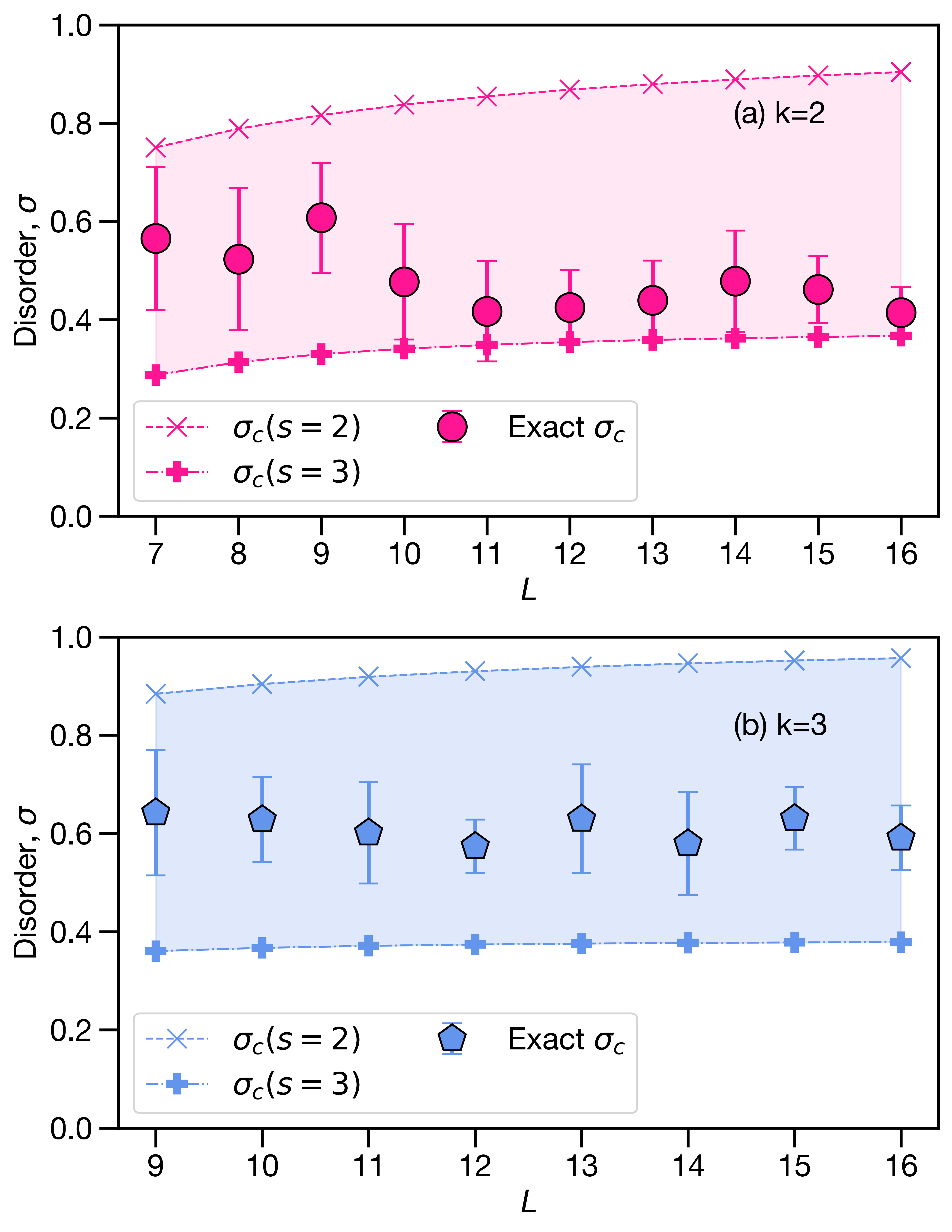}
    \caption{Critical disorder strength $\sigma_{c}$, is plotted as a function of system size $L$ for $k=2$ in panel (a) and $k=3$ in panel (b). The exact critical disorder threshold is identified at the point where the ground state enters into the bulk. Each data point represents an average over $10$ independent realizations of the Hamiltonian coefficients with error bars representing one standard deviation. The shaded dashed region indicates the critical disorder range obtained using Eq.\eqref{eq:sigma_c_for_a_general_s} for $s=2$ and $s=3$.}
    \label{fig:critical_disorder_exact_two_sigma_three_sigma}
\end{figure}
Figure \ref{fig:critical_disorder_exact_two_sigma_three_sigma} shows that the average exact critical disorder always lies between the critical disorder estimates using Eq. \eqref{eq:sigma_c_for_a_general_s}, when we consider two $(s = 2)$ and three $(s = 3)$ standard deviations from the mean. Thus, we conclude that the exact critical disorder scales as
\begin{eqnarray}
    \sigma_{c} \sim | \mu|.
\end{eqnarray}

\textit{An intermediate regime} -- Another regime where universal features may emerge from this model is known as the double-scaling limit, in which $\alpha \equiv k^{2} / L$ is held fixed \cite{Berkooz2019, Cotler2017}. In this limit, the density of states of the Hamiltonian $H^{(L, k)}(0, 1)$ can be calculated analytically \cite{Erdős2014},  exhibiting a sharp cutoff at $[-R, R]$, where $R = 2/ \sqrt{1 - e^{- 4 \alpha/ 3}}$. We leave this as a direction for future research.

In general, the critical disorder can roughly be viewed as the strength of disorder at which the disordered part of the Hamiltonian becomes comparable to the deterministic part. From Eq. \eqref{eq:normalized_single_parameter_hamiltonian}, we see that the disorder-free and the disordered part scale roughly as ($|| A||$ denotes the spectral norm of the matrix $A$)
\begin{eqnarray}
    \bigg|\bigg| \frac{\mu L H^{(L, k)}(1,0)}{\sqrt{3^{k}} \binom{L}{k}} \bigg|\bigg| & \sim & |\mu| L \;, \\
    \bigg|\bigg| \frac{\sigma L H^{(L, k)}( 0, 1)}{\sqrt{3^{k} \binom{L}{k}}} \bigg|\bigg| & \sim &  |\sigma| L \;,
\end{eqnarray}
the ratio of these two terms gives the critical disorder $\sigma_{c} \approx  | \mu|$.

\section{Conclusions and outlook}\label{eq:conclusions}
In this article, we first classified the $k-$local Hamiltonians based on the locality $k$ and the system size $L$. We showed that based on whether the system size and the locality are even or odd, these Hamiltonians with $\mu = 0$ belong to the three canonical random matrix ensembles -- GOE, GUE, and GSE. We then showed that the Hamiltonians with $\mu \neq 0$ can be interpreted as deformed random matrices. This analogy allowed us to analytically derive that energy gap between the ground state and first excited state vanishes for $\mu < 0$. We found that for locality $k \gg \sqrt{L}$ the gap closes at $\sigma_{c} \approx |\mu|$, and for locality $k \ll \sqrt{L}$ we showed that the critical disorder scales with $|\mu|$ without a sharp cutoff. Additionally, for a large locality, we showed that the energy gap has a universal quadratic scaling. Whether such a universal scaling exists for smaller locality $(k = 2, 3)$ or in the double scaling limit $(k^{2} / N = $ constant) remains an open question.

While our work has focused on all-to-all interacting Hamiltonians as abstract models, it is natural to ask whether any of these models, particularly for small locality $k = 2, 3$ can be realized in experiments. The long-range nature of the interactions suggests potential implementation in cavity QED platforms \cite{Mivehvar02012021, Douglas2015}. For small values of $k$ these Hamiltonians remain within the reach of present-day experimental platforms \cite{Katz2023, Isenhower2011, PRXQuantum.4.030311, Liu2008, PhysRevA.61.012302}. 

Beyond experimental realization, the models studied here can serve as a toy or semi-solvable example for exploring universal properties of quantum chaos and spectral gaps. Their all-to-all connectivity allows analytical control over the eigenstructures and the spectral scaling while retaining nontrivial random matrix behavior. This combination of analytical control and complexity makes them ideal for probing questions that are otherwise intractable in short and long-range Hamiltonians in general. The Sachdev–Ye–Kitaev (SYK) model serves as a notable reference for how analytically tractable disordered systems can capture aspects of chaotic dynamics, quantum gravity, and holography \cite{RevModPhys.94.035004}.

Several promising directions remain open for future research. First, while our analysis assumes all-to-all interactions, in most experimental platforms implementing large $k$ is not physical. Extending these results to systems with restricted connectivity—such as nearest-neighbor or power-law interactions—will clarify the robustness of the observed classification and gap behavior under realistic constraints.  Another key direction is to examine the robustness of the gapped phase under symmetry-breaking perturbations to the Hamiltonian. The robustness of spectral gaps under various local perturbations of the Hamiltonian remains an active area of research \cite{DeRoeck2019, Nachtergaele2024, Michalakis2013, 10.1063/1.3490195}.

In its most general form, the $k$-local Hamiltonian lacks the symmetries that are often present in physically realistic models. To better reflect such systems, it is important to explore the impact of imposing symmetries on the Hamiltonian. Introducing symmetry constraints may reduce the number of terms in the Hamiltonian, or put constraints on the coefficients -- resulting in a sparser structure compared to the fully general case. Some works have been done in the local interaction and translation invariant case \cite{PhysRevLett.119.220504, PhysRevB.100.035113}. Beyond these results, several open questions remain, particularly regarding physical realizability and robustness under perturbations. These research directions are left open for future investigation.

\begin{acknowledgments}
The author thanks Michael Kolodrubetz, Vedika Khemani, Matteo Ippoliti, Samuel Begg and Yves Kwan for useful discussions. The author also thanks the anonymous referee for their constructive comments and suggestions. This work was carried out with the support from the National Science Foundation (NSF) under Grant No. OSI-2228725. The author thanks the High Performance Computing facility at The University of Texas at Dallas (HPC@UTD) for providing computational resources. 
\end{acknowledgments}

\section*{V. DATA AVAILABILITY}
All data that support the findings of this article are openly available ~\cite{dowarah2025zenodo}.

\appendix
\section{Degeneracy of the ground state for even $L$ and even $k$ in the thermodynamic limit} \label{sec:degeneracy_of_largest_eigenvalue_even_L_even_k}
In this section, we show that when both $L$ and $k$ are even, then the ground state is degenerate in the thermodynamic limit, although the degeneracy appears to be lifted for finite $L$. To simplify the notation, we denote $H^{(L, k)}(1, 0) \equiv H_{0}$ and $H^{(L, k)}(0, 1) \equiv V$. Here we present the proof for $k = 2$, the generalization to higher even values of $k$ follows straightforwardly. We begin by defining the operators
\begin{eqnarray}
    \tau_{z} &=& \frac{1}{\sqrt{3}} (\sigma_{x} + \sigma_{y} + \sigma_{z}) \label{eq:tau_z} \\
    U &=& \begin{bmatrix}
        \langle 0 | u_{+}\rangle & \langle 0|u_{-}\rangle \\
        \langle 1|u_{+}\rangle & \langle 1 |u_{+}\rangle
\end{bmatrix}
\end{eqnarray}
where the states $|u_{-}\rangle$ and $|u_{+}\rangle$ are defined in Eq.\eqref{eq:eigenstates_of_sigma_x+sigma_y+sigma_z}. We further define the following
\begin{eqnarray}
    \tau_{x} &=& U \sigma_{x} U^{\dagger}, \\
    \tau_{y} &=& U \sigma_{y} U^{\dagger}.
\end{eqnarray}
These operators act on the eigenstates as
\begin{eqnarray}
    \tau_{z} |u_{\pm}\rangle &=& \mp |u_{\pm}\rangle, \\
    \tau_{x} |u_{\pm}\rangle &=& |u_{\mp}\rangle, \\
    \tau_{y} |u_{\pm}\rangle &=& \mp i|u_{\mp}\rangle.
\end{eqnarray}
These new operators obey the standard commutation relation of Pauli matrices: $[\tau_{a}, \tau_{b}] = 2 i \epsilon_{abc} \tau_{c}$. Thus there exists a real orthogonal matrix $R$ such that $\tau_{\alpha} = \sum_{\alpha} R_{\mu \alpha} \sigma_{\alpha}$. The matrix $R$ represents a rotation of $2 \pi /3$ that maps the $z$-axis to the $(1, 1, 1)/\sqrt{3}$ direction. In this new basis, the completely disordered matrix $V$ retains the same structure
\begin{eqnarray}
    V &=& \sum_{j_{1} < j_{2}} \Tilde{J}_{j_{1} j_{2}} \tau^{(j_{1})}_{\alpha_{1}} \tau^{(j_{2})}_{\alpha_{2}}
\end{eqnarray}
with the new coupling constants $\Tilde{J}_{j_{1} j_{2}}$ having the same mean and variance as the original coupling constants.

Consider the degenerate subspace spanned by the orthonormal eigenstates of $H_{0}$
\begin{eqnarray}
    |\psi_{+} \rangle = |+ + \cdots +\rangle, \; 
    |\psi_{-} \rangle = |- - \cdots -\rangle,
\end{eqnarray}
with an equal unperturbed energy $E_{0} = \mu L$. We then add the perturbation
\begin{eqnarray}
    H &\equiv & \frac{\mu L}{\sqrt{3^{k}} \binom{L}{k}}H_{0} + \sigma \frac{L}{\sqrt{3^{k} \binom{L}{k}}} V
\end{eqnarray}
Let us set $\mu = 1$ and treat $\sigma$ as a perturbation parameter ($\sigma < 1)$. Then using the projectors
\begin{eqnarray}
    P &=& |\psi_{+}\rangle \langle \psi_{+}| + |\psi_{-}\rangle \langle \psi_{-}|, \\
    Q &=& I - P,
\end{eqnarray}
the effective Hamiltonian within the degenerate subspace can be written as \cite{Sakurai_Napolitano2020, BRAVYI20112793}
\begin{eqnarray}
    H_{\mathrm{eff}} &=& E_{0} P + \sigma P V P \\
    && \; + \sigma^{2} P V Q \frac{1}{E_{0} - H_{0}} Q VP + \cdots \\
    &=& E_{0} P \\
    && \; + \sum_{r > 0} \sigma^{r} P V (Q (E_{0} - H_{0})^{-1} V)^{r - 1} P \qquad
\end{eqnarray}
The off-diagonal elements that lift the degeneracy are
\begin{eqnarray}
    \langle \psi_{+}| H_{\mathrm{eff}}|\psi_{-}\rangle &=& \sum_{r} \sigma^{r} \langle \psi_{+}|V (Q (E_{0} - H_{0})^{-1} V)^{r - 1}|\psi_{-}\rangle \nonumber
\end{eqnarray}
Now consider the action of $V$ on the states $|u_{+}\rangle$, $|u_{-}\rangle$. It is easy to verify that a typical $2-$local term of them form $\tau_{\alpha_{1}} \otimes \tau_{\alpha_{2}}$ can flip at most two spins: $|u_{-}\rangle \to \pm |u_{+}\rangle$ or $|u_{+}\rangle \to \pm|u_{-}\rangle$. Thus starting from $|\psi_{-} \rangle = |---\cdots -\rangle$ it will require $r = L/2$ ($L$ is even) successive applications of $V$ to reach $|\psi_{+} \rangle = |+++ \cdots +\rangle$. Therefore, for all $r < L/2$
\begin{eqnarray}
    \langle +++ \cdots +| H_{\mathrm{eff}}| --- \cdots -\rangle &=& 0
\end{eqnarray}
Hence, the first nonzero off-diagonal element appears at order $r = L/2$, giving an exponentially small splitting $\mathcal{O}(\sigma^{L/2})$. Since $\sigma < 1$, this splitting vanishes in the limit $L \to \infty$, implying that the two states are degenerate in the thermodynamic limit.

\section{Critical disorder estimation for $k \ll \sqrt{L}$}
\label{sec:critical_disorder_scaling_k_2_derivation}
In this section, we derive the critical disorder scaling for $k=2$ in the large $L$ limit. Let us start with Eq.\eqref{eq:sigma_c_for_a_general_s}
\begin{eqnarray}
    s L \sqrt{\sigma^{2}_{s} + \mu^{2}/ \binom{L}{k}} &=& \mu L \\
    && \hspace{1 mm} + \frac{\sigma^{2}_{s} L^{2}}{2^{L}} \sum_{v \neq \mu L } \frac{1}{\mu L- v}. \label{eq:k_2_s_sigma_c}
\end{eqnarray}
Define
\begin{eqnarray}
     B &\equiv& \frac{1}{2^{L}} \sum_{v \neq \mu L} \frac{1}{\mu L - v} = \frac{1}{2^{L}} \sum_{n\neq 0, L} \binom{L}{n} \frac{1}{\mu L - \lambda^{(L,k)}_{n}}. \nonumber
\end{eqnarray}
The factor $\binom{L}{n}$ appears due to the degeneracy of the eigenvalues corresponding to each $n$. Then in this notation, Eq.\eqref{eq:k_2_s_sigma_c} becomes
\begin{eqnarray}
   \mu + B \sigma^{2}_{s} L &=& s \sqrt{\sigma^{2}_{s} + \mu^{2} / \binom{L}{k}}.
\end{eqnarray}
Squaring both sides
\begin{eqnarray}
    B^{2} L^{2} \sigma^{4}_{s} + (2 \mu B L - s^{2}) \sigma^{2}_{s} + (\mu^{2} - s^{2} \mu^{2} / \binom{L}{k}) &=& 0. \nonumber
\end{eqnarray}
Using the quadratic formula, we obtain
\begin{eqnarray}
    \sigma^{2}_{s} &=& \frac{1}{2 B^{2} L^{2}} ( s^{2} - 2 \mu B L \pm \\
    && s \sqrt{s^{2} - 4 \mu B L - 4 \alpha^{2} B^{2} L^{2}/ \binom{L}{k}^{-1}} ). \label{eq:quadratic_roots_for_critical_disorder}
\end{eqnarray}
For $k=2$, from Eq. \eqref{eq:analytical_eigenvalues_of_disorder_free_Hamiltonian}, we obtain
\begin{eqnarray}
    \lambda^{(L,2)}_{n} &=& \mu L \binom{L}{2}^{- 1} \sum^{2}_{j = 0} (-1)^{j} \binom{n}{j} \binom{L - n}{2 - j},
\end{eqnarray}
which in the large $L$ limit becomes
\begin{eqnarray}
    \lambda^{(L,2)}_{n} &\approx & \frac{2 \mu}{L} (\binom{n}{0} \binom{L - n}{2} - \binom{n}{1} \binom{L - n}{1} \\
    && \quad + \binom{n}{2} \binom{L - n}{0}), \\
    &=& \mu (L - 4 n - 1 + 4 n^{2}/L).
\end{eqnarray}
Using this, $B$ in the large $L$ limit becomes
\begin{eqnarray}
    B &\approx& \frac{1}{2^{L} \mu} \sum^{L- 1}_{n=1} \binom{L}{n} \frac{1}{4n + 1 - 4n^{2}/L}.
\end{eqnarray}
To simplify $B$, let us define $y=(2n - L)/\sqrt{L}$, which gives $n=L/2 + \sqrt{L}y/2$, and $\Delta y = 2/\sqrt{L}$. Then using De Moivre-Laplace theorem \cite{Feller1968}
\begin{eqnarray}
    \frac{1}{2^{L}} \binom{L}{n} &\approx & \sqrt{\frac{2}{L}} \frac{1}{\sqrt{\pi}} e^{-\frac{1}{2}(\frac{2n-L}{\sqrt{L}})^{2}}, \\
    &=& \frac{1}{\sqrt{2 \pi}} e^{ - y^{2}/2} \Delta y.
\end{eqnarray}
Using this, we can write the sum as
\begin{eqnarray}
    B &\sim & \frac{1}{\mu} \int^{\sqrt{L}}_{-\sqrt{L}} e^{- y^{2}/2} \frac{1}{L + 1 - y^{2}} dy.
\end{eqnarray}
Using the saddle point method \cite{ARFKEN2013551} around $y=0$, we find that the integral scales as
\begin{eqnarray}
    B \sim 1 / (\mu L) \; .
\end{eqnarray}
Then, substituting the scaling $ B \sim 1/(\mu L)$ in Eq. \eqref{eq:quadratic_roots_for_critical_disorder}, we obtain
\begin{eqnarray}
    \sigma^{2}_{s} \sim \frac{\mu^{2}}{2} (s^{2} - 2 \pm |s| \sqrt{s^{2} - 4 + 4 \binom{L}{k}^{-1}}).
\end{eqnarray}
From this equation, we conclude that irrespective of the value of $s=2, 3$ considered,
critical disorder scales as
\begin{eqnarray}
    \sigma_{c} \sim |\mu|\; .
\end{eqnarray}
\section{Eigenvalues and eigenstates of the disorder-free Hamiltonian}
In this section, we derive the eigenvalues and eigenstates of the disorder-free Hamiltonian $H^{(L, k)}(1, 0)$. We start by defining the operator
$P^{(j)} = (\sigma^{(j)}_{x} + \sigma^{(j)}_{y} + \sigma^{(j)}_{z}) / \sqrt{3}$,
with the properties
\begin{eqnarray}
    P^{(j_{1})} P^{(j_{2})} &=& P^{(j_{2})} P^{(j_{1})}, \\
    (P^{(j)})^{2} &=& I.
\end{eqnarray}
Let $|u_{\pm}\rangle$ be the eigenstates of $P$. Now for a given $s=(s_{1}, s_{2}, \cdots, s_{L})$, where $s_{j} \in \{-1, 1\}$, we define the state
\begin{eqnarray}
    |\Psi_{s}\rangle &=& \bigotimes^{L}_{j=1} |u_{s_{j}} \rangle,
\end{eqnarray}
which satisfies
\begin{eqnarray}
    P^{(j)} |\Psi_{s}\rangle &=& s_{j} |\Psi_{s}\rangle.
\end{eqnarray}
The $k-$local disorder-free Hamiltonian satisfies
\begin{eqnarray}
    H^{(L,k)}(1, 0) &=& \sum_{|J| = k} \prod_{j \in J} P^{(j)}.
\end{eqnarray}
Define the function
\begin{eqnarray}
    G(t) &=& \prod^{L}_{j=1} (I + t P^{(j)}) = \sum^{L}_{k=0} t^{k} H^{(L,k)}(1, 0),
\end{eqnarray}
whose action on the state $|\Psi_{s} \rangle$ is given by
\begin{eqnarray}
    G(t) |\Psi_{s}\rangle &=& \prod^{L}_{j=1} (1 + t s_{j}) |\Psi_{s} \rangle, \\
    &=& (1 + t)^{N_{+}} (1 - t)^{N_{-}} |\Psi_{s} \rangle, \\
    &=& \sum^{L}_{k=0} \lambda^{(L, k)}(s) t^{k} |\Psi_{s} \rangle,
\end{eqnarray}
where $N_{\pm} = \# \{ j: s_{j} = \pm 1\}$, and $\lambda^{(L, k)}$ is the coefficient of $t^{k}$ in $(1 + t)^{N_{+}} (1 - t)^{N_{-}}$. Comparing the coefficient of $t^{k}$ on both sides, we obtain
\begin{eqnarray}
    H^{(L, k)} |\Psi_{s} \rangle &=& \lambda^{(L, k)}(s) |\Psi_{s} \rangle.
\end{eqnarray}
Using the binomial theorem, we obtain the eigenvalues
\begin{eqnarray}
    \lambda^{(L, k)}(s) &=& \sum^{k}_{j=0} (-1)^{j} \binom{N_{-}}{j} \binom{N_{+}}{k - j}.
\end{eqnarray}
\vspace{1 cm}
\bibliography{apssamp}
\end{document}